# Bacterial stress granule protects mRNA through ribonucleases exclusion

Linsen Pei[1,2,8], Yujia Xian[1,2,8], Xiaodan Yan[1,2,8], Charley Schaefer[4,8], Aisha H. Syeda[4,5,8], Jamieson A. L. Howard[4], Hebin Liao[1,2], Fan Bai[6,7]*, Mark C. Leake[4,5]*, Yingying Pu[1,2,3]*

[1]The State Key Laboratory Breeding Base of Basic Science of Stomatology & Key Laboratory of Oral Biomedicine Ministry of Education, School & Hospital of Stomatology, Medical Research Institute, Wuhan University, China

[2]Frontier Science Centre for Immunology and Metabolism, Wuhan University, China

[3]Department of Immunology, Hubei Province Key Laboratory of Allergy and Immunology, State Key Laboratory of Virology and Medical Research Institute, Wuhan University School of Basic Medical Sciences, China

[4]School of Physics, Engineering and Technology, University of York, York, UK.

[5]Department of Biology, University of York, York, UK.

[6]Biomedical Pioneering Innovation Centre (BIOPIC), School of Life Sciences, Peking University, Beijing, China.

[7]Beijing Advanced Innovation Center for Genomics (ICG), Peking University, Beijing, China.

[8]These authors contributed equally to this work.

*Corresponding authors. Email: yingyingpu@whu.edu.cn; fbai@pku.edu.cn; mark.leake@york.ac.uk

Membraneless droplets formed through liquid-liquid phase separation (LLPS) play a crucial role in mRNA storage, enabling organisms to swiftly respond to environmental changes. However, the mechanisms underlying mRNA integration and protection within droplets remain unclear. Here, we unravel the role of bacterial aggresomes as stress granules (SGs) in safeguarding mRNA during stress. We discovered that upon stress onset, mobile mRNA molecules selectively incorporate into individual proteinaceous SGs based on length-dependent enthalpic gain over entropic loss. As stress prolongs, SGs undergo compaction facilitated by stronger non-specific RNA-protein interactions, thereby promoting recruitment of shorter RNA chains. Remarkably, mRNA ribonucleases are repelled from bacterial SGs, due to the influence of protein surface charge. This exclusion mechanism ensures the integrity and preservation of mRNA within SGs during stress conditions, explaining how mRNA can be stored and protected from degradation. Following stress removal, SGs facilitate mRNA translation, thereby enhancing cell fitness in changing environments. These droplets maintain mRNA physiological activity during storage, making them an intriguing new candidate for mRNA therapeutics manufacturing.

In genetics, messenger RNA (mRNA) plays a crucial role as an intermediary, transferring genetic information from DNA to ribosomes for protein synthesis. Despite its transient nature, recent investigations have shown that it is possible to preserve mRNA for extended periods. Non-membrane bound messenger ribonucleoprotein particles (RNPs)[1], formed through LLPS, may play a role in mRNA storage and enable cell stress response, diversity, and development[2]. During periods of cellular stress, mRNA molecules that are not immediately necessary for survival are confined in stress granules (SGs)[3,4]. RNP droplets are also involved in RNA storage and translation regulation in germ cells, both spermatocytes and oocytes, ensuring only necessary mRNA molecules are translated during embryonic development[5-7]. Nevertheless, the impact of eukaryotic RNP granules in RNA translation and degradation remains under debate, and some studies have reported that SGs have no effect on RNA translation or stability[8,9].

In our previous studies, we identified the bacterial aggresomes as proteinaceous, membraneless LLPS droplets that form in response to a wide range of cellular stress [10,11]. Here, we reveal for the first time aggresomes as the SGs in bacteria, serving to protect mRNA in stress conditions. Remarkably, this discovery mirrors a notable phenomenon observed in eukaryotic cells, wherein mRNAs undergo translational repression prior to entering SGs, only to resume translation upon SG disassembly[4]. The correlation between mRNA length, SG size, and mRNA translation status observed in eukaryotic systems finds resonance in our investigations of bacterial SGs[4]. Moreover, our elucidation of the mechanisms underpinning mRNA protection within these droplets represents a significant advancement. This protective function aids in cell recovery following stress removal, thereby promoting enhanced fitness and facilitating antibiotic tolerance.

## Results

**Aggresomes are bacterial SGs that are enriched with mRNA**

ATP depletion is a crucial factor inducing aggresome formation[10,11]. Arsenite is traditionally known as an oxidative stressor causing ATP decrease in eukaryotic cells[12], our experiments also demonstrate its effectiveness in inducing cellular ATP depletion in bacterial cells (Extended Data Fig. 1a). By exposing *Escherichia coli* (*E. coli*) cells to 2 mM arsenite, we were able to consistently induce aggresome formation within just 30 minutes. This rapid induction enabled a more expedited exploration of the dynamics of aggresome formation compared to previously used stress factors[10,11], as visualized by distinct foci of green fluorescent protein (GFP) fused aggresome biomarker HslU (HslU-GFP) in cells (Fig. 1a). To investigate the presence of RNA along with proteins in bacterial aggresomes, we purified aggresomes from the lysed cells using immunoprecipitation with an antibody specific to HslU-GFP (Fig. 1b). Purified aggresomes exhibited a generally spherical shape and remained intact throughout the process of lysis and purification (Fig. 1c). Subsequently, RNA purification was performed, yielding 1.64 ± 0.11 fg (±SD) RNA per aggresome (Fig. 2f).

We then performed transcriptome analysis of aggresome RNA (Aggr-RNA) and RNA which remained in the cytosol (Cyto-RNA). Pairwise correlation analysis (Extended Data Fig. 1b) and differential gene expression analysis (Fig. 1d and Extended Data Fig. 1c, d) revealed that Aggr-RNA was distinct from Cyto-RNA. Kyoto Encyclopedia of Genes and Genomes (KEGG) pathway analysis revealed that aggresomes were significantly enriched with mRNA critical for cell growth and proliferation, including transcripts involved in the tricarboxylic acid (TCA) cycle, general carbon metabolism, and the biosynthesis of peptidoglycan and secondary metabolites (Fig. 1e). In contrast, ribosomal RNA (rRNA) was found to be depleted from aggresomes (Fig. 1f). These findings, along with the observation of aggresome formation through phase separation[9] and the identification of proteins and RNAs associated with stalled translation initiation complexes in aggresomes (Extended Data Table S1), suggest that bacterial aggresomes exhibit similar characteristics to eukaryotic stress granules (SGs). Therefore, we propose to categorize them as bacterial SGs.

We selected transcripts based on the RNA-seq and qPCR analysis, which indicated significantly different expression levels between aggresomes and cytosol following arsenite induction (Extended Data Fig. 1e). The first group of transcripts included *nuoJ, nlpE, glpK* and *mukF,* which presented high enrichment scores in aggresomes. The second group included *dps, gmhA* and *gppA,* which exhibited high enrichment scores in the cytosol. We individually labelled these mRNA with Pepper RNA aptamer[13] in a background strain genomically expressing HslU-mCherry and visualized with HBC530 green dye under time-resolved structured illumination microscopy (SIM) (Extended Data Fig. 2a-c). In the absence of arsenite, no HslU-mCherry foci formation was observed, and all transcripts were uniformly distributed throughout the cytosol (Extended Data Fig. 2d, Videos S1-S4). However, after arsenite induction, *nuoJ, nlpE, glpK* and *mukF* transcripts gradually formed distinct foci, which colocalized with HslU-mCherry foci (Fig. 1g-j; Videos S5-S8), with the rate of foci formation variable across different transcripts (Extended Data Fig. 2e-h). Consistent findings were observed with HokB induction and under conditions of starvation (Fig. 1c; Extended Data Fig. 3a-d); both scenarios led to cellular ATP depletion[10, 14] (Extended Data Fig. 1f). These results illustrate a specificity of transcriptomes in aggresomes. In contrast, transcripts of *dps*，*gmhA* and *gppA* remain uniformly distributed throughout the cytosol after arsenite treatment (Extended Data Fig. 3e; Videos S9-S11).

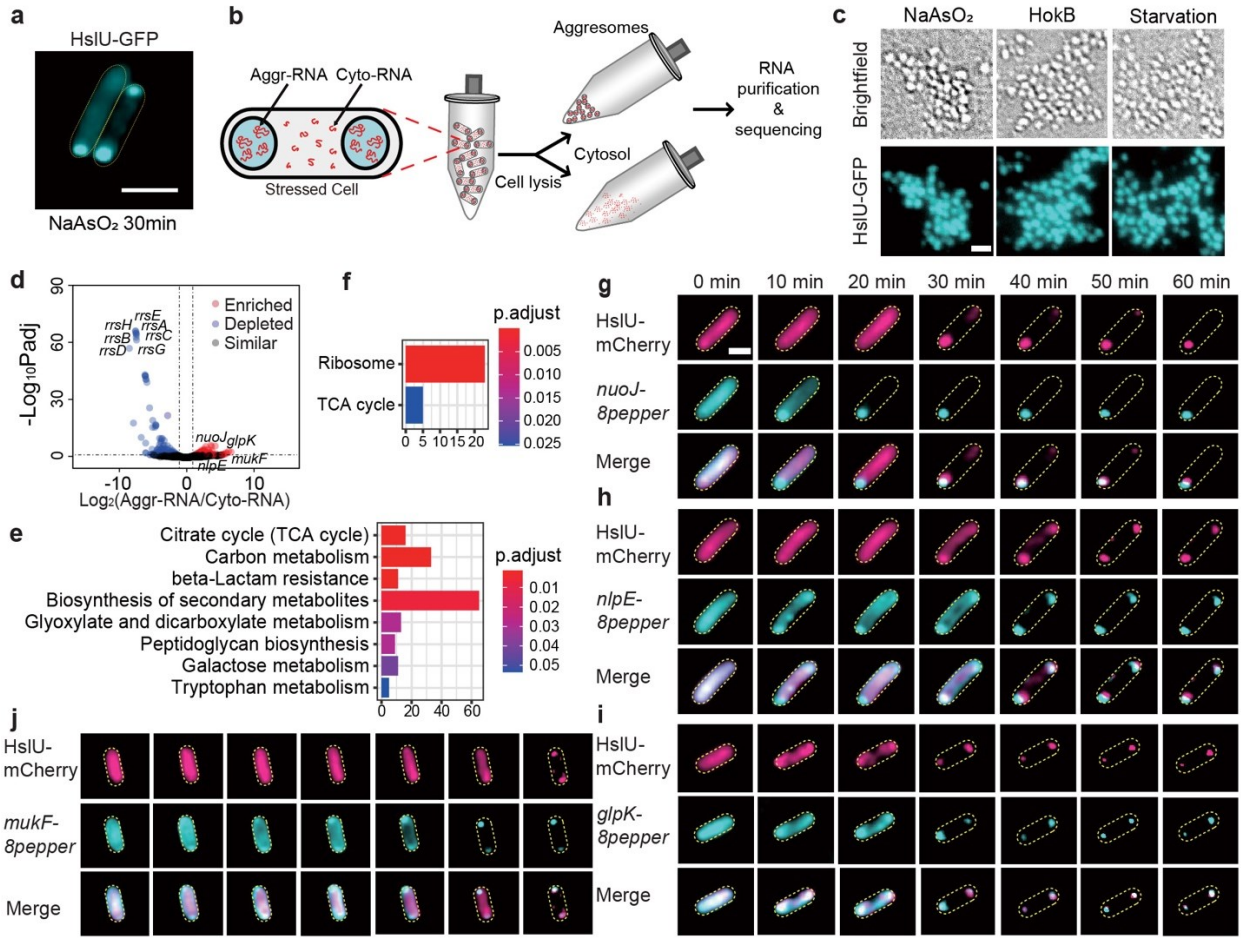

Fig. 1 | Aggresome formation requires mRNA. **a.** Formation of distinct HslU-GFP foci (cyan) in cells after 30min of 2mM NaAsO$_2$ exposure, cell body indicated (yellow dash), scale bar: 2μm. **b**. Schematic summary of aggresome isolation process for RNA-seq, Aggr-RNA, aggresome RNA; Cyto-RNA, cytosol RNA. **c**. Purified aggresomes from *E. coli* cells treated with 2mM arsenite (NaAsO$_2$), HokB overexpression (HokB) and starvation, respectively, scale bar: 1 μm. **d**. Volcano plot depicting RNA abundance (Deseq2) purified from aggresomes (Aggr-RNA) versus purified from aggresome depleted cytosol (Cyto-RNA). Red dots indicate RNAs that are significantly enriched in aggresomes (log$_2$ fold change (Aggresome-RNA/Cytosol-RNA) >1; adjusted p value < 0.01). Blue dots indicate RNAs that are significantly depleted in aggresomes (log$_2$ fold change (Aggresome-RNA/Cytosol-RNA) < -1; adjusted p value < 0.01). Black dots indicate RNAs that are either not significantly enriched or fail to meet the fold change requirement. **e**. KEGG pathway analysis of Aggr-RNA. **f**. KEGG pathway analysis of Cyto-RNA. **g-j**. Time-lapse imaging showing that mRNAs of *nuoJ* (**g**), *nlpE* (**h**), *glpK* (**i**) and *mukF* (**j**) partition into aggresomes under arsenite treatment. Dual-colour microscopy of HslU labeled with mCherry (magenta) and mRNA-8pepper labeled by HBC530 (cyan), scale bar, 1 μm.

**Prolonged stress triggers bacterial SGs compaction**

Quantitative transcriptome analysis, using operonic mRNA as a reference, revealed that mRNA molecules enriched in aggresomes have an average length that exceeds five times the length of that in cytosol (Extended Data Fig. 4a, b). High sensitivity RNA ScreenTape analysis validated these findings with arsenite induction (p value < 0.0005, Fig. 2a and Extended Data Fig. 4c). Similar observations were also observed using other stress induction factors (Extended Data Fig. 4d, e).

The observed influence of RNA length on its incorporation into aggresomes aligns with recent studies on LLPS droplets of transcriptional machinery in mammalian cells, where stress granule-enriched mRNAs are often longer than the average length[15]. To provide a theoretical framework for our experimental observations, we adopted Flory-Huggins (FH) theory for ternary mixtures[16] where the non-specific interactions between the protein, RNA and solvent are captured by three FH interaction parameters (Methods and Supplementary Information). We first considered the regime of zero or very low aggresomal RNA concentrations in which the formation and compaction of droplets (Fig. 2c) has been driven by the protein-water interaction parameter. In this regime, the model assumes that RNA has no direct influence on the phase separation process itself. In the steady-state regime of the model, the partitioning of RNA (Fig. 2b) is simply driven by an RNA-protein interaction that is favorable compared to the RNA-water interaction, which is enhanced by high protein concentrations in compact droplets. Here, the extent of RNA partitioning only depends on a single emergent parameter, which is modelled as a consensus enthalpy parameter per nucleotide base, $\Delta H$ (see Equation 11 in Supplementary Information). Therefore, the total enthalpy per RNA molecule is predicted to simply increases linearly with its length due to an increased number of nucleotide-protein interactions. We find a reasonable fit to the experimental data spanning two orders of magnitude for the ScreenTape intensity ratio and RNA length using an optimized enthalpy parameter of $\Delta H = (5.5 \pm 0.7) \times 10^{-3}\ k_BT$ per nucleotide base (Fig. 2b). This prediction quantitatively agrees with our experimental data, providing a comprehensive understanding of the distribution of RNA lengths inside an aggresome at steady-state, through the lens of statistical thermodynamics.

Our model, based on a mass balance analysis of a closed system, makes predictions about the effects of prolonged stress on aggresome dynamics, which indicate that prolonged stress leads to decreased compatibility between proteins and solvents, consequently prompting the compaction of aggresomes. As aggresome compaction sets in, the thermal energy scale increases, thereby resulting in an increased affinity for RNA molecules (Fig. 2c, and Equations 10,11 in Supplementary Information). Notably, the increased affinity for RNA molecules during aggresome compaction is predicted in the model to facilitate increased transfer of cytosolic mRNA to the aggresomes.

To experimentally test this prediction, we extended arsenite treatment from 30 to 60 minutes and employed SIM and transmission electron microscopy (TEM) to measure the aggresome area as projected on the camera detector. The results confirmed the model's prediction, showing an approximately 50% decrease in the measured cross-sectional area of aggresomes when stress is prolonged. This reduction in size was observed for both the aggresome biomarker protein HslU and the Pepper-labeled aggresomal mRNA (Fig. 2d, e, Extended Data Fig. 5a-h). Furthermore, to quantify the mass amount of aggresomal RNA present, we used flow cytometry to determine the absolute number of aggresomes in a given volume and subsequently performed RNA purification. The results show that prolonged arsenite stress leads to an approximate 40% increase in the mass of RNA per aggresome (Fig. 2f) and an approximate 140% increase in the mass of protein per aggresome (Extended Data Fig. 5r). These findings are consistent with similar observations made during aggresome induction using other stress factors (Extended Data Fig. 5i-p). These experimental results validate the steady-state model prediction that prolonged stress triggers aggresome compaction, characterized by a reduction in size concurrent with an increase in mass.

**There is dynamic flux of RNA between aggresomes and the cytosol**

Meanwhile, we observed a discernible decrease in the mean length of aggresomal mRNA transcripts (Fig. 2g). We hypothesize that the aggresome provides an environment in which RNA is protected from enzymatic degradation mechanisms that occur in the cytosol. To account for this effect, we adapted the steady-state model to include the potential effects of dynamic RNA mobility in the cytosol, and RNA flux between the cytosol and aggresome. Here, the adapted model comprises dynamic coupled processes which involve diffusion of cytosolic RNA, the incorporation of RNA into an aggresome, and then the potential exit of the RNA from the aggresome (for full details see Supplementary Information). In this dynamic framework, although the strength of enthalpic binding of RNA in the aggresome is higher for longer RNA, as incorporated into the steady-state model, the rate of RNA diffusion within the cytosol is slower for longer RNA, described by $K_{diff,C} = K^0_{diff} N_R^{-1/2}$, where $N_R$ is the RNA length in terms of number of nucleotide bases. However, after an RNA molecule is incorporated into an aggresome it may then exit with a probability which decreases with increasing RNA length since it is proportional to the Boltzmann factor $\exp(-N_R \Delta H/k_B T)$. When RNA is in the cytosol, the dynamic model assumes it may undergo enzymatic degradation, described as a process whereby each bond between nucleotides may break with a rate $K_{deg}$. The dynamic interplay between diffusion and degradation then leads to a net depletion of RNA from the aggresome over time. This depletion process begins with shorter RNA chains, as they have a higher probability of exiting the aggresome. Some of these shorter RNA fragments may then diffuse back into the aggresome, creating a dynamic equilibrium. These processes are governed by master equations (see Equations 15 and 16-17 in the Supplementary Information), which describe the time evolution of RNA concentrations in the aggresome and cytosol. These coupled dynamic processes result in a complex interplay between diffusion, degradation, and protection within the aggresome, ultimately shaping the RNA length distribution over time. A clear result of this dynamic RNA flux is that at later times following extended arsenite treatment the dynamic model predicts a decrease in the mean RNA length inside the aggresome, even though all the RNA degradation *per se* has occurred *outside* the aggresome in the cytosol. A further prediction of the model is that there is typically a much greater decrease in the mean RNA length in the cytosol over comparable time scales.

To test these dynamic model predictions, we quantified aggresome-RNA, cytosol-RNA, and total-RNA from the cells subjected to different stresses for varying durations (Fig. 2i-k, Extended Data Figs. 5l, 5p, 6). Our results demonstrate an increase in the mass of Aggresome-RNA and a decrease in cytosol-RNA during prolonged stresses, confirming the transfer of RNA from cytosol to aggresomes as predicted by the model. Meanwhile, total-RNA decreased during prolonged stress, suggesting that degradation occurred. Based on this analysis, we infer that the depletion of cytosol-RNA results from both transfer and degradation. We note also that there has been evidence suggesting similar dynamic activity of RNA in eukaryotic stress[17,18], albeit indirectly from studying the dynamic flux of SG proteins. Building upon this foundation, our bacterial study breaks new ground by presenting the first direct evidence, to our knowledge, of mRNA dynamics in any type of stress granule. This finding not only parallels the observations in eukaryotic systems but also extends our understanding of RNA behavior in prokaryotic stress responses.

To determine the impact of cellular stress on the mobility of aggresomal mRNA, we employed high-sensitivity millisecond single-molecule Slimfield microscopy[15] to track its spatial localization inside living cells (Extended Data Table S2, Video S12[19] and Extended Data Fig. 5q). By photobleaching the majority of labeled mRNA molecules in live cells, we could track individual molecules and determine their apparent diffusion coefficient from the initial gradient of the mean square displacement relative to tracking time. After 30 minutes incubation with arsenite, the mean diffusion coefficient of the four aggresome mRNA transcripts ranged from $0.43 \pm 0.06$ μm$^2$/s (±SE; number of tracks $N = 159$) for *nuoJ* to $0.84 \pm 0.12$ μm$^2$/s ($N = 129$) for *mukF* (Fig. 2h and Extended Data Table S2). These values were significantly higher than the mean diffusion coefficient of the aggresome protein biomarker HslU (approximately 0.2 μm$^2$/s), consistent with single mRNA molecules exhibiting liquid-like diffusion within aggresomes. Prolonged incubation for 60 minutes with arsenite resulted in a mean diffusion coefficient ranging from $0.36 \pm 0.06$ μm$^2$/s ($N = 147$) for *nuoJ* to $0.64 \pm 0.13$ μm$^2$/s ($N = 49$) for *glpK*. These findings suggest a trend towards a more viscous state within the aggresome during prolonged stress, affirming stress-induced compaction characterized by a simultaneous reduction in size and an increase in mass.

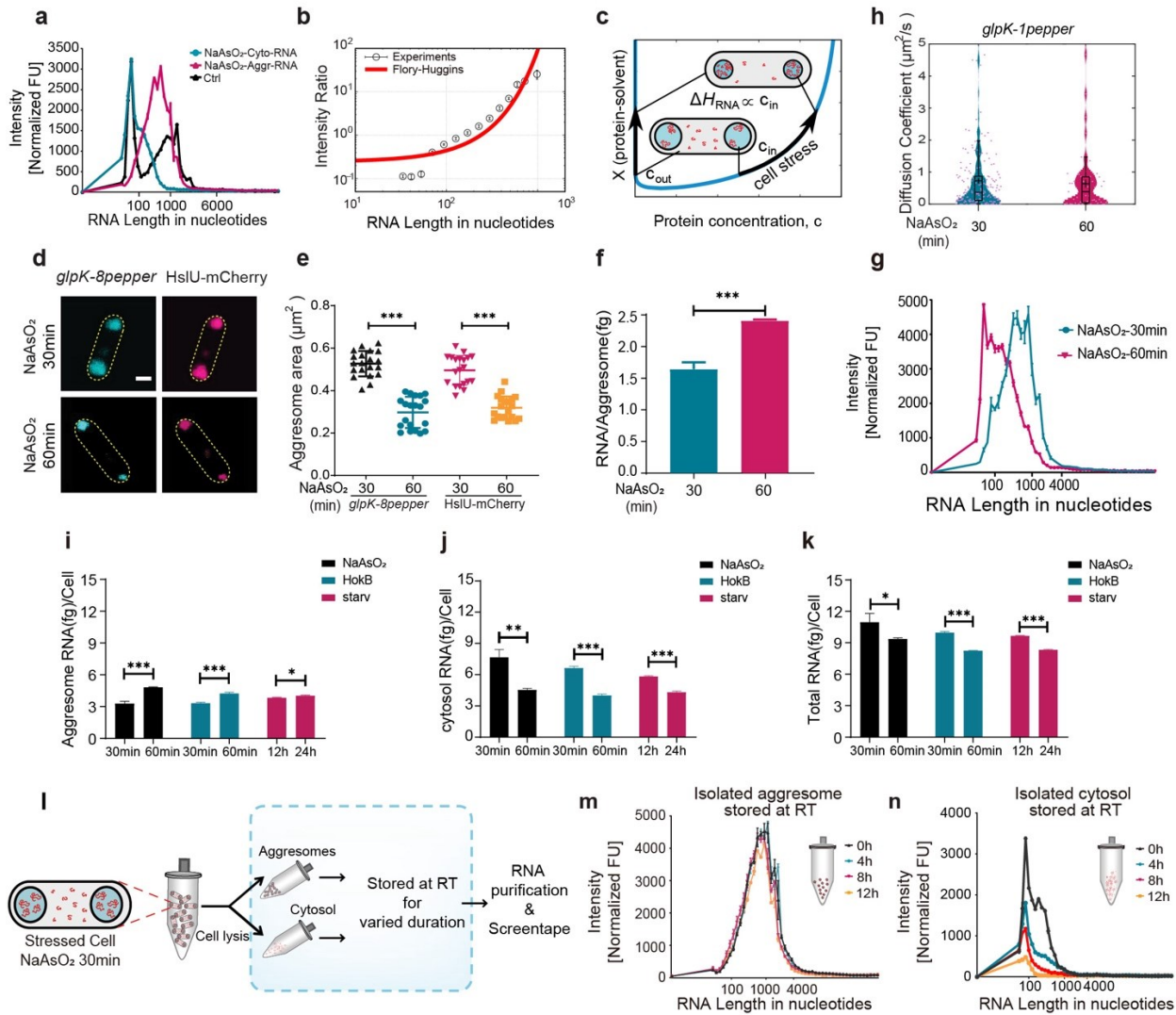

Fig. 2 | Prolonged stress triggers aggresome compaction and RNA dynamics in aggresomes. a. Distribution of RNA length in nucleotide determined by ScreenTape RNA analysis, NaAsO$_2$-Aggr-RNA: aggresome RNA from the cells with arsenite (NaAsO$_2$) treatment; NaAsO$_2$-Cyto-RNA: cytosol RNA from the cells with arsenite treatment; control (Ctrl): total RNAs from the cells of exponential phase. b. Aggresome/Cytoplasm RNA intensity ratio versus RNA length (SD error bars), with overlaid fit (red) based on Flory-Huggin's theory with optimized parameters Prefac = 0.32 ± 0.08 and ΔH =0.0055 ± 0.0007 k$_B$T, goodness-of-fit R$_2$=0.989378. c. The Flory-Huggins phase diagram for the solvent and protein predicts droplet compaction. The blue curves represent the binodal concentrations in the cytoplasm and in the aggresome. An increasing protein-solvent interaction parameter, χ, (interpreted to correlate with cell stress) leads to an increasing binodal protein concentration inside the aggresome; due to mass conservation this causes the droplets to shrink. d. Representative SIM images of cells with different duration of NaAsO$_2$ treatment, *glpK-8pepper* represents *glpK-8pepper* RNA stained with HBC530 green dye, HslU-mCherry represents protein HslU labeled by fluorescent protein mCherry, Scale bar, 500 nm. e. Statistical analysis of the area of aggresomes projected onto the camera detector in cells with different duration of NaAsO$_2$ treatment, based on SIM images. f. The mass of RNA per aggresome in cells with different duration of NaAsO$_2$ treatment, the total mass of RNA in bulk aggresomes (RNA$_{total}$) is measured by a Qubit fluorometer, the number of aggresomes (N$_{aggresome}$) of a given volume was determined by FACS, and RNA/Aggresome is determined by RNA$_{total}$/N$_{aggresome}$(n=3). g. Distribution of RNA length in aggresomes determined by ScreenTape RNA analysis, NaAsO$_2$-30min indicates aggresomes from cells treated with arsenite for 30min, NaAsO$_2$-60min indicates aggresomes from cells treated with arsenite for 60min. h. Distribution of diffusion coefficients as violin jitter plot for *glpK* transcript in aggresomes with arsenite exposure for 30 minutes or 60 minutes. i. The mass of RNA in aggresomes per cell subjected to different stresses for varying durations. j. The mass of RNA

in cytosol per cell subjected to different stresses for varying durations. **k**. The mass of total RNA per cell subjected to different stresses for varying durations. The total mass of RNA in different cell compartments is measured by a Qubit fluorometer, the number of cells in a given volume was determined by a colony counting assay. **l**. Schematic summary of aggresome and cytosol isolation and storage processes for subsequent RNA purification and ScreenTape analysis. **m**. Isolated aggresomes protecting RNA *in vitro*. RNAs were extracted from isolated aggresomes that have been stored at room temperature for 0, 4, 8, and 12 hours, respectively. Isolated aggresomes derived from cells treated with arsenite for 30 minutes. **n**. RNA undergoing degradation in isolated cytosol *in vitro*. RNAs were extracted from isolated cytosol that have been stored at room temperature for 0, 4, 8, and 12 hours, respectively. Isolated cytosol derived from cells treated with arsenite for 30 minutes. Unpaired Student's t-test was performed, and the error bar indicates SE; *$P < 0.05$, **$P < 0.005$, and ***$P < 0.0005$.

**Aggresomes protect mRNA by exclusion of mRNA ribonucleases**

To further explore whether aggresomes play a role in protecting sequestered mRNA, we extracted aggresomes and cytosols from arsenite stressed cells. Subsequently, these isolated aggresomes and cytosols were stored at room temperature for varying durations (0, 4, 8, and 12 hours. Following this, RNA was purified from the aggresomes, cytosols for subsequent ScreenTape analysis (Fig. 2l). Notably, our results revealed a striking preservation of RNA transcriptome within aggresomes over prolonged periods at room temperature (Fig. 2m). Conversely, RNA in the cytosol of stressed cells exhibited a gradual degradation pattern over time (Fig. 2n). These live-cell and *in vitro* findings together suggest that aggresomes play a pivotal role in safeguarding RNA integrity. In addition, we conducted quantitative polymerase chain reaction (qPCR) assays to compare the degradation rates of individual mRNA transcripts in aggresomes and cytosol. We used *talB* and *gltI* as test transcripts because RNA-seq indicates that they have equivalent enrichment levels between aggresomes and cytosol after 30 minutes of arsenite treatment. We designed probes to target both 3' and 5' ends as well as a short region in the middle of each transcript. The qPCR data indicated comparable levels of relative expression for each probe region of each transcript in both aggresomes and the cytosol after 30 minutes of arsenite treatment (Extended Data Fig. 7a,c). However, after prolonged stress for 3 hours, the relative levels of both transcripts were significantly reduced in the cytosol compared to the levels in the aggresome (Extended Data Fig. 7b,d), confirming that aggresomes play a role in protecting sequestered mRNA.

To investigate the mechanism of mRNA protection in aggresomes, we used affinity purification mass spectrometry (AP-MS) to explore proteomes enriched or depleted in aggresomes. AP-MS analysis output in the Search Tool for the Retrieval of Interacting Genes/Proteins (STRING) database revealed distinct protein-protein interaction (PPI) network topologies for aggresomes compared to the cytosol, with mean clustering coefficients of $0.673 \pm 0.033$ and $0.496 \pm 0.01$, respectively (Extended Data Fig. 7e-g). Of particular significance, ribonucleases specialized for rRNA processing (RNG and RIA) and for tRNA processing (RnpA) were found to be enriched in aggresomes, whereas RNAses responsible for mRNA degradation, including Exoribonuclease II (RNB), Ribonuclease BN (RBN) and Oligoribonuclease (ORN), were depleted from aggresomes (Fig. 3a). SIM imaging, using the HslU-GFP aggresome biomarker, confirmed that as aggresomes form and mature following arsenite induction, mCherry-fused RNB, RBN and ORN under the control of their own native promoters, were excluded from aggresomes throughout the entire process (Fig. 3d-f). Similar results were observed using other stress factors (Extended Data Fig. 7h-m). Taken together, these data indicate that exclusion of mRNA ribonucleases plays a potential role in facilitating the protection of aggresomal mRNA.

**Protein surface charge plays a key role in exclusion of mRNA ribonucleases**

Our statistical thermodynamics steady-state model, based on Florey-Huggins theory, provided an explanation for the exclusion of RNases from aggresome droplets. In this model, we incorporated different interaction parameters to account for the higher affinity of ribonucleases to the water solvent and their repulsion away from the aggresome. Using bioinformatics analysis, we revealed a significant net negative surface charge that distinguishes RNA nucleases (RNB, -20; RBN, -9; ORN, -10) from most RNA binding proteins (RBPs) present within aggresomes (Fig. 3b-c and Extended Data Fig. 8a). Building on these findings, we hypothesized that the exclusion of ribonucleases from aggresomes is influenced by negative charge repulsion, possibly from the RNA molecules, which have a high net negative charge due to the phosphate backbone, already present within the condensate.

To test this hypothesis, we conducted mutagenesis by substituting all aspartic acid (D) or glutamic acid (E) residues across the entire enzyme center with alanine (A), which resulted in a drastic change in surface charge (RNB-ME, 82; RBN-ME, 26; ORN-ME, 20) (Extended Data Fig. 8b) while also disrupting the RNA binding motif and catalytic centre. In this scenario, the mutated ribonucleases were able to enter aggresomes without influencing aggresome RNA condensation. As anticipated, SIM imaging following arsenite induction clearly showed distinct foci of mCherry-fused RNB-ME, RBN-ME, and ORN-ME, which colocalized effectively with HslU-GFP foci (Fig. 3g-i). These compelling results support the notion that the surface charge plays a pivotal role in the exclusion of mRNA ribonucleases from aggresomes. Furthermore, RNB plays a significant role in mRNA degradation, which enzymatically hydrolyzes single-stranded polyribonucleotides in a processive 3' to 5' direction[20]. We conducted mutagenesis by substituting only the D and E residues located outside the RNA binding motif and catalytic center of RNB with an A residue (RNB-MO). This resulted in a moderate positive surface charge (Extended Data Fig. 8c), while the protein's enzymatic activity remained intact (Extended Data Fig. 8d). In this context, RNB-MO was able to enter aggresomes and effectively degrade mRNA within them. As anticipated, SIM imaging following arsenite induction demonstrated a depletion of RNA from aggresomes, with positively charged RNB-MO entering (Fig. 3j-m, Extended Data Fig. 9). RNA ScreenTape analysis confirmed these results (Fig. 3n). These findings support a mechanism involving the interplay between attractive protein-RNA interactions, electrostatic repulsion between RNAases and RNA, and length-dependent entropic differences between the aggresome and cytosol environment, for how mRNA is preserved within LLPS droplets.

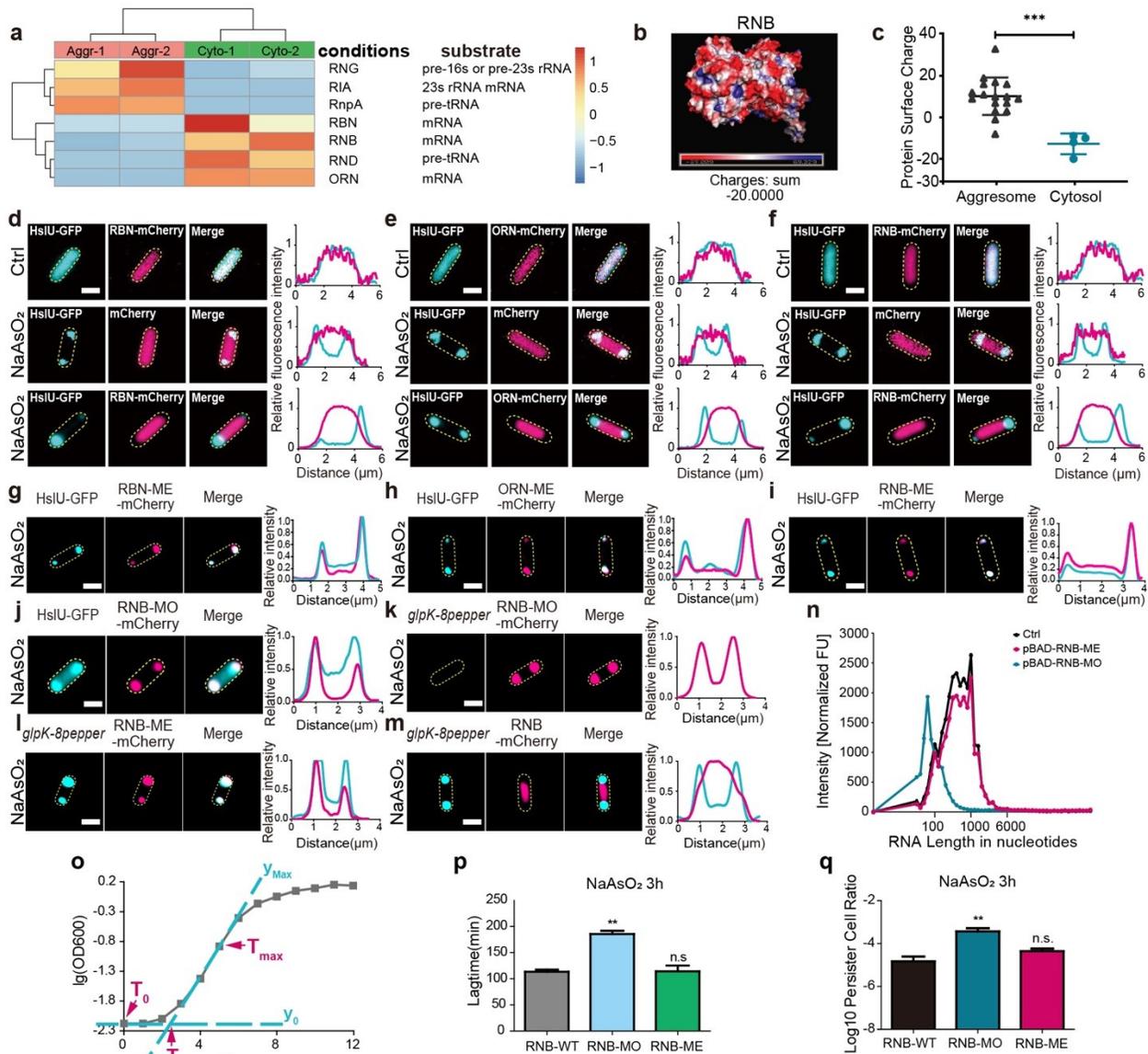

Fig. 3 | Aggresome protects mRNA by excluding mRNA ribonucleases. **a**. Heatmap showing relative protein abundance of RNases in aggresomes (Aggr) or in cytosol (Cyto). Scale beside the heatmap indicates $log_2$-normalized transcript abundance relative to the mean expression level. **b**. Protein surface charge of ribonuclease RNB. **c**. Statistical analysis of protein surface charges of RNA binding proteins in aggresomes or that of RNases in cytosol. **d-j**. Distribution of ribonucleases or their mutant proteins, RBN (**d**), ORN (**e**), RNB (**f**), RBN-ME (**g**), ORN-ME (**h**), RNB-ME (**i**), RNB-MO (**j**) in cells with chromosomally labeled HslU-EGFP under arsenite treatment. Each ribonuclease or the mutant is labeled by mCherry. **k-m**. Distribution of RNB-MO (**k**), RNB-ME (**l**), RNB (**m**) in cells with *glpK-8pepper*/HBC530, each ribonuclease or the mutant is labeled by mCherry. ME: Mutation across the Entire enzymatic center; MO: Mutation Outside the regions of RNA binding motif and catalytic center. Relative fluorescence intensities of GFP and mCherry along the long axis of each cell are measured and plotted (chart on the right of each row of fluorescent images). Ctrl, control, indicating exponential phase cells without drug treatment; Ars, arsenite treatment. **n**. ScreenTape analysis of aggresome RNA from cells overexpression RNB-ME or RNB-MO after 30 min of 2 mM arsenite treatment (n=3). Ctrl, aggresome RNA from wild-type cells after 30 min of 2 mM arsenite treatment (n=3). **o**. Method of determining lag time. The specific method involves identifying the starting point ($T_0$) and the point of maximum growth rate ($T_{max}$) on the growth curve, and then constructing the tangent line equations $Y_0$ and $Y_{max}$ for these two points, respectively. The intersection of these two tangent line equations corresponds to the time point where the lag time occurs ($T_{Lag}$) (Buchanan et al, 1990). **p**. Lag time of

strains containing wild-type RNB or RNB-MO and RNB-ME after exposure to 2 mM arsenite for 3 hours. **q**. Persister cell ratio in strains containing wild-type RNB or the mutant RNB-MO and RNB-ME after exposure to 2 mM arsenite for 3 hours.  Scale bar, 1 μm. Unpaired Student's t-test was performed, and the error bar indicates SE; *P <0.05, **P < 0.005, and ***P < 0.0005.

**Protection of aggresome mRNA increases cell survival following stress**

Time-resolved SIM showed that, following removal of arsenite stress, the transcripts of *glpK* (Fig. 4a) and *nuoJ, nlpE, mukF* (Extended Data Fig. 10a-d) were released from aggresomes into the cytosol over a period of 1-3 hours. Previous mass spectrometry and RNA-seq analysis showed that both mRNA and protein products of *glpK* accumulated in aggresomes at high levels compared to the cytosol during stress (Extended Data Fig. 10e, f). To assess the potential recycling of aggresome released mRNA for translation, we used an arabinose-inducible fluorescent reporter of the GlpK protein. Following arsenite induction, GlpK-GFP localized to aggresomes as expected. We then photobleached the entire cell and removed arsenite from the imaging buffer. To prevent the influence of newly transcribed mRNA, we introduced sub-Minimum Inhibitory Concentration (sub-MIC) rifampicin to inhibit RNA synthesis. Additionally, to account for potential pre-existing fluorescent proteins maturing after photobleaching, we supplemented the medium with tetracycline, which inhibits bacterial protein synthesis by targeting the 30S ribosomal subunit. Consequently, in cells treated with rifampicin alone, the reappearance of GlpK-GFP fluorescence can be attributed to both new synthesis of the protein from aggresome-released RNA and pre-existing fluorescent proteins maturing after photobleaching. In contrast, in cells treated with both rifampicin and tetracycline, the reappearance of GlpK-GFP fluorescence can be attributed solely to pre-existing fluorescent proteins maturing after photobleaching. Importantly, the reappearance of the fluorescence in cells treated with rifampicin alone is significantly higher than that in cells treated with both rifampicin and tetracycline (Fig. 4b, c), thereby confirming the release of preserved mRNA from aggresomes for subsequent protein synthesis.

These results underscore the importance of stored RNA in facilitating a faster response to fluctuating environmental conditions and enhancing cell fitness. Furthermore, we hypothesized that the depletion of RNA from aggresomes could prolong the lag time required for cellular recovery. Our findings revealed that strains carrying the mutant RNB-MO exhibited longer lag times following stress exposure compared to those harboring the wild-type RNB (Fig. 3o, p; Extended Data Fig. 11a, b). Interestingly, this prolonged lag time correlated with an increase in the overall fraction of persister cells within the populations (Fig. 3q; Extended Data Fig. 11c,d). These observations suggest a potential link between RNA availability and cellular recovery kinetics.

To determine if mRNA protection by aggresomes confers survival advantages, we performed screening of small chemical compounds (Fig. 4d; Extended Data Fig. 12), identifying conditions of media supplementation with 1,6-hexanediol (Hex) or doxorubicin (Dox). Hex is commonly used to study biomolecular LLPS, since it dissolves liquid condensates but not solid-like aggregates via the disruption of RNA-protein interactions, while Dox is a nucleic acid intercalator which inhibits RNA gelation *in vitro* and dissolves RNA nuclear foci in eukaryotic cells[21,22]. Hex or Dox alone has no effects on inducing aggresome formation or promoting persister formation (Extended Data Fig. 13). However, aggresome formation was impaired in cells treated with Hex or Dox after arsenite exposure (Fig. 4d, e). We compared the survival difference between untreated cells, and those treated with Hex or Dox during arsenite induction or other cell stress factors. Untreated cells, which exhibited expected aggresome formation, consistently showed a significantly higher persister ratio after different types of antibiotic treatment. On the contrary, cells treated with Hex or Dox showed significantly reduced aggresome formation correlated to a lower prevalence of persisters (Fig. 4f, g). Taken together, these observations indicate that aggresomes enable protection of captured mRNA from ribonuclease degradation, which facilitates translation of released mRNA following stress removal, resulting in a significantly increased cell fitness during stress conditions.

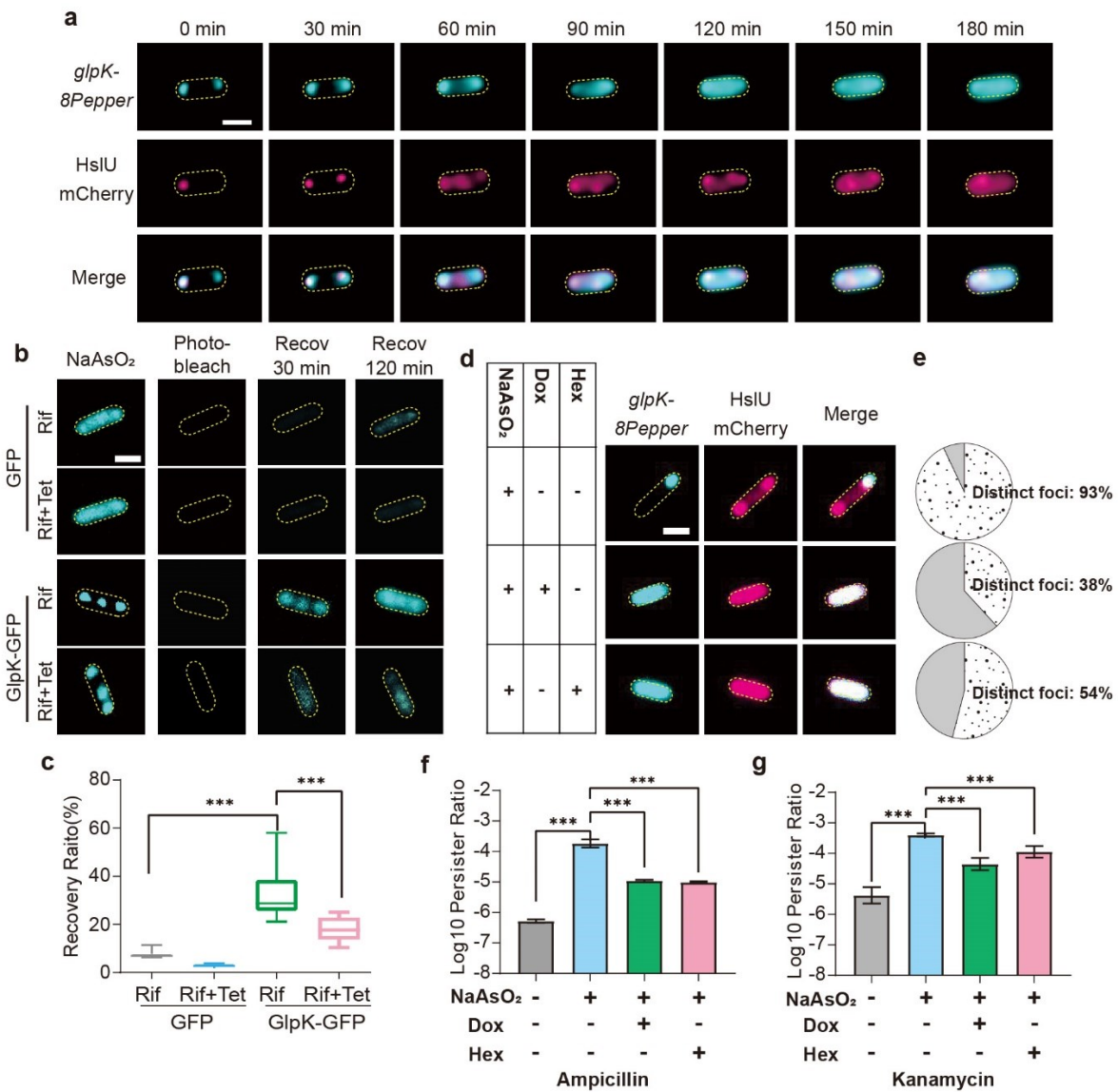

Fig. 4 | The translation of released mRNA from aggresomes promotes cell fitness. **a**. Releasing of *glpK-8pepper* mRNA from aggresomes into cytoplasm upon arsenite removal. **b**. Translation of *gfp* or *glpK-gfp* mRNA after aggresome disassembly. Rif, rifampicin; Tet, tetracycline; Recov, recovery time. **c**. Statistical analysis of GFP fluorescence recovery ratio at 120 min after photobleaching, depicted from (**b**). **d**. Representative fluorescence images of cells with different chemical combination treatment. Ars, arsenite; Dox, doxorubicin; Hex, 1,6-hexanediol. e. Statistical analysis of cells with distinct mRNA foci after different chemical combination treatment, data from (**d**) (n=100). **f-g**. Cell survival rate (log scale) after 4 hours (**f**) ampicillin (**g**) kanamycin killing (n=3). Scale bar, 1 μm. Unpaired Student's t-test was performed, and the error bars indicate SE; n=3, *P <0.05, **P < 0.005, and ***P < 0.0005.

**Discussion**

Aggresomes are bacterial stress granules that store mRNA under stress conditions. These specialized structures offer several advantages that make them a model system for studying RNA storage mechanisms under stress. One notable advantage is that their formation through LLPS can be finely tuned, allowing for a well-regulated and organized assembly of RNP droplets. Furthermore, the relative stability of aggresomes permits efficient purification, while preserving the integrity of the aggresome transcriptome and proteome. These attributes make bacterial aggresomes ideally suited for the investigation of how RNP droplets integrate and safeguard mRNA, shedding light on essential cellular processes during stress responses.

Drawing upon experimental advantages and polymer physics modeling, our research has uncovered essential dynamics within aggresomes during stress conditions. Initially, upon stress onset, long transcripts, potentially forming multivalent RNA-protein interactions, are preferentially incorporated into aggresomes. However, when stress is prolonged, the protein-solvent compatibility decreases and the aggresomes undergo compaction, resulting in smaller droplets but with higher density that in turn facilitate stronger non-specific interactions between the proteins and RNA. Consequently, the compact droplets promote recruitment of increasingly short RNA chains, indicating a dynamic and adaptive process within the aggresomes. Remarkably, our findings demonstrate that mRNA ribonucleases are effectively repelled from aggresomes, ensuring the integrity and preservation of mRNA within these structures under stress conditions. Our mutagenesis studies offer valuable insights into the potential origin of these exclusion forces, specifically indicating a significant trend towards a net negative surface charge for the excluded ribonucleases. Given that RNA molecules themselves possess a net negative charge, electrostatic repulsion from the aggresome likely contributes to this exclusion effect. Additionally, the high surface electrical polarity of the ribonucleases enhances their enthalpic attraction to water, further reinforcing their exclusion from aggresomes. In support of our statistical thermodynamics model, our subsequent mutagenesis experiments have provided confirmatory evidence for the role of surface charge in the exclusion of mRNA ribonucleases.

Collectively, our discoveries offer mechanistic insights into how droplets shield vital mRNA molecules during stress, thereby advancing our understanding of biomolecular condensates and their functional roles in cellular physiology. While some of our findings align with observations in *C. crescentus*, such as the storage of mRNA within biomolecular condensates, others diverge significantly. Notably, our discovery that ribonucleases responsible for mRNA degradation are excluded from stress granules contrasts with findings in *C. crescentus*, where the ribonuclease RNaseE is integral to the formation of BR bodies[23]. Instead, the organization of BR bodies appears to modulate degradosome activity[24, 25]. Furthermore, these findings hold promise for informing future strategies aimed at developing tailored synthetic aggresomes for therapeutic applications involving mRNA, such as vaccines and gene silencing.

## AUTHOR CONTRIBUTIONS

Conceptualization: Y.P., F.B., M.L.; Methodology: L.P., Y.X., X.Y., C.S., A.S., J.H., H.L.; Investigation: L.P., Y.X., X.D., C.S., A.S., J.H.; Bioinformatics analysis: X.Y.; Supervision: Y.P., M.L.; Writing – original draft: Y.P., M.L.; Writing – review & editing: Y.P., M.L., F.B.

## ACKNOWLEDGMENTS

This work was supported by the grants from the National Key R&D Program of China (2021YFC2701602), the National Natural Science Foundation of China (31970089, T2125002), Science Fund for Distinguished Young Scholars of Hubei Province (2022CFA077), the Fundamental Research Funds for the Central Universities (2042022dx0003), the Engineering and Physical Science Research Council (EP/W024063/1, EP/Y000501/1) and Biotechnology and Biological Sciences Research Council (BB/W000555/). We also thank all the staff in the Core Facilities of Medical Research Institute at Wuhan University and the Core Facilities at School of Life Sciences at Peking University for their technical support.

## DECLARATION OF INTERESTS

Authors declare that they have no competing interests.

## CODE AVAILABILITY STATEMENT

Bioinformatic scripts that was used to process the data in this study is available on GitHub, https://github.com/123456yxd/Code-of-RNA-seq.

## DATA AVAILABILITY STATEMENT

Sequencing data are available under Gene Expression Omnibus (GEO) accession number GSE243619. Other data that was used in this study is available on Zenodo, DOI: 10.5281/zenodo.12755876 and 10.5281/zenodo.12758836.

# Methods

**Bacterial strains and plasmids construction**. Generation of strains with a chromosomal gene-fluorescent protein translational fusion were performed using λ-red mediated recombination system following previous established protocols[26]. Relative fluorescent protein fragment (GFP and mCherry) was inserted to replace the stop cassette of the selection-counter-selection template plasmid (pSCS3V31C) as previously described[27]. Next, the GFP-Toxin-CmR or mCherry-Toxin-CmR fragment was amplified from the template plasmid and then transformed into electrocompetent cells, along with an induced recombineering helper plasmid (pSIM6), as described in previously[28]. The transformed cells were plated on selection plates containing relative antibiotics. The Toxin-CmR cassette was subsequently removed from the chromosome via another round of λ-red mediated recombination, utilizing a counter selection template. Finally, the cells were plated on counter selection plates containing rhamnose to activate the toxin.

To construct strains with an overexpression plasmid or plasmids with native promoter of target proteins, the bacterial recombinant protein vector pBAD and p15G was used in this study, respectively. The target protein fragment was first amplified from wild type *E. coli* MG1655, recombined with the pBAD or p15G plasmid via homologous recombination and then transformed into electrocompetent DH5α cells. The transformed cells were plated on selection plates containing relative antibiotics. Correct recombinant plasmids were subsequently extracted using the FastPure Plasmid Mini Kit (Vazyme DC201-01) and the plasmids were transformed into relative electrocompetent cells according to the experimental requirements. All the plasmids used in this study were listed in Extended Data Table S3 and the primers for strains and plasmids construction were listed in Extended Data Table S4.

**RNA labeling.** To perform RNA labeling experiments, an mRNA-pepper recombinant plasmid which involved fusing an 8-pepper tag to the 3' end of the target mRNA was first constructed. Overnight bacterial cultures expressing pepper-mRNA were diluted by 1:100 into fresh LB and incubated in a shaker (220 rpm) for 2 hr. Next, the expression of the target mRNA was induced by adding 0.001% arabinose (w/v) and incubated in a shaker (220 rpm) for additional 4 hr. The treated cells were then collected by centrifugation at 4000g for 5 mins and suspended with Imaging Buffer (40 mM HEPES, pH 7.4, 100 mM KCl, 5 mM $MgCl_2$ buffer, 1:100 HBC ligands). Cells were stained for 5 mins, protected from light at room temperature.

**Brightfield and fluorescence microscopy.** Brightfield and fluorescence imaging were performed on a Nikon Ti2-E inverted fluorescence microscope. Illumination was provided by a solid-state laser at wavelength 488 nm for GFP. The fluorescence emission signal of cells was imaged onto a pco.edge camera.

**SIM microscopy.** To perform SIM super-resolution microscopy, overnight bacterial cultures were diluted 1:100 into fresh LB medium and incubated in a shaker at 220 rpm for 2 hr. To induce the expression of the target protein, 0.001% arabinose (w/v) was added, and samples were further incubated in a shaker at 220 rpm for 2 hr. Imaging was carried out using a Nikon N-SIM S microscope, with illumination provided by solid-state lasers at wavelengths of 405 nm for Tag-BFP and Hoechst, 488 nm for GFP and 561 nm for mCherry.

**Time-resolved widefield imaging.** The Flow Cell System FCS2 (Bioptechs) system was used to perform the time-resolved widefield imaging[4]. Treated bacterial cultures were harvested by centrifugation, washed, suspended in PBS buffer and then imaged on a gel-pad made up of 3% low gelling temperature agarose, with a cell culture to gel-pad volume ratio of 1:10. The gel-pad was positioned at the center of the FCS2 chamber and surrounded by LB liquid medium buffer containing 20 mM $NaAsO_2$. The cells were observed under bright-field/epifluorescence illumination at 37 ℃, with images captured every five mins over a period of 180 mins. Imaging was carried out using a Nikon N-SIM S microscope.

**Aggresome purification and extraction of aggresome RNA.** MG1655 bacterial cultures expressing HslU-GFP were incubated in a shaker at 220 rpm and 37 °C until they reached the exponential phase and then treated with 2 mM $NaAsO_2$ for 30 mins as previously described[5]. The cells were collected by centrifugation at 6,500 rpm for 5 mins and washed with PBS buffer three times. The resulting pellets were resuspended in 1 mL of lysis buffer (50 mM Tris-HCl pH 7.4, 100 mM KOAc, 2 mM MgOAc, 0.5 mM DTT, 50 μg/mL Heparin, 0.5% NP40, complete mini EDTA-free protease inhibitor (Bimake, B14001), and 1 U/μL of RNasin Plus RNase Inhibitor (Vazyme, R301-03). The cells were lysed by repeated freeze-thaw method[29], and the lysates were then centrifuged at 1000 x g for 5 mins at 4°C to remove cell debris. The isolation of total and aggresome RNA was achieved by transferring 50 μL and 950 μL of the supernatants, respectively, to new microcentrifuge tubes, followed by extraction using the Bacteria RNA Extraction Kit (Vazyme, R403-01). The 950 μL supernatant was then centrifuged at 18,000g for 20 mins at 4 °C to enrich the aggresome. The resulting supernatant was discarded, and the pellet was washed with 1 mL of lysis buffer and then resuspended in 300 μL of lysis buffer. The enriched aggresome was pre-washed twice with equilibrated DEPC-treated Protein A Dynabeads, which were then removed using a magnet. The aggresome was then affinity purified using 20 μL of anti-GFP antibody (MBL,048-3) as previous described[30]. The solution was then centrifuged at 18,000g for 20 mins at 4 °C, and the supernatant was discarded to remove any unbound antibody. The pellet was then resuspended in 500μL of lysis buffer and 60 μL of equilibrated DEPC-treated Protein A Dynabeads. The sample was rotated for 3 hr at 4 °C. The Dynabeads were washed three times with wash buffer 1, once with wash buffer 2 and once with wash buffer 3. The beads were then resuspended in 200μL of 100μg/mL Protease K solution and incubated for 15 mins at 37°C. Trizol reagent (Vazyme, R403-01) was added to the samples, and RNA was extracted following the manufacturer's protocol.

**Quantification of Aggresomal, Cytosolic and Total Cellular RNA Content.** After stress stimulation as indicated, a 100 mL bacterial culture was subjected to plating for CFU counting to estimate the total number of cells. The cells were then lysed for the extraction of total RNA. The total RNA content was quantified by Qubit 3.0, then divided by the total number of cells to calculate the total RNA content per individual cell. Aggresomes were isolated and purified from the cell lysate and the total number was measured using flow cytometry. RNA was extracted from all isolated aggresomes, and the RNA content per aggresome was calculated by dividing the total aggresomal RNA content by the number of aggresomes. The cytosolic RNA content per cell was obtained by subtracting the total RNA content of all aggresomes from the total cellular RNA content and dividing the result by the total number of cells. This process was repeated three times (n=3) to ensure reproducibility.

**RNA Concentration Measurements Using a Qubit Fluorometer.** The Qubit Fluorometer 3 (Q33216, Invitrogen), in conjunction with the Equalbit RNA HS Assay Kit (EQ211, Vazyme), was employed for accurate RNA concentration measurements. The working solution was prepared and aliquoted, with 10 μL of Standard

#1 and 10 μL of Standard #2 each diluted into 190 μL of the working solution for machine calibration and the generation of an accurate standard curve. Given the excellent linear relationship of the assay kit within the range of 5 ng-100 ng/μL for RNA, RNA samples were diluted prior to measurement to ensure their concentrations fell within this range. For each sample, 1 μL of RNA was mixed with 199 μL of the working solution, and the mixture was loaded into the Qubit Fluorometer for automatic concentration determination based on the standard curve.

**RNA-seq.** The total RNA of the samples was extracted using the Bacteria RNA Extraction Kit (R403-01, Vazyme) and subjected to mRNA selection, fragmentation, cDNA synthesis, and library preparation using the VAHTSTM Total RNA-seq (H/M/R) Library Prep Kit for Illumina® (NR603, Vazyme). The library quality was analyzed on a Bioanalyzer. High-throughput sequencing was conducted on the Genome Analyzer IIx (Illumina).

**RT-qPCR.** To prepare the samples for RT-qPCR analysis, overnight cultures were 1:100 diluted into fresh LB medium and incubated in a shaker at 220 rpm for 3 hr, then treated with 2 mM $NaAsO_2$ for varying lengths of time according to the experimental requirements. After treatment, the cells were harvested by centrifugation and total RNA was extracted using the Bacteria RNA Extraction Kit (Vazyme China). To synthesize cDNA, 300 ng of RNA was used with the HiScript III RT SuperMix for qPCR (+gDNA wiper) (Vazyme China), following the manufacturer's instructions. The qPCR reactions were performed using the ChamQ Universal SYBR qPCR Master Mix (Vazyme China) and the BIO-RAD CFX Connect Real-Time PCR Detection System from Bio-Rad (USA), according to the manufacturer's instructions. The primers for RT-qPCR analysis were listed in Extended Table S4. The relative amount of mRNA is determined by normalizing it to the level of the internal control gene, 16s RNA.

**ScreenTape assay.** To conduct the ScreenTape experiment, the High Sensitivity RNA ScreenTape assay kit (Agilent, America) were utilized to achieve high accuracy in measuring RNA samples at a concentration of approximately 10 ng/μL. Prior to analysis, all RNA samples were diluted, and 1 μL of RNA buffer was thoroughly mixed with 2 μL of RNA, followed by incubation at 72 °C for 3 mins and 1 min on ice. Subsequently, individual samples were subjected to RNA length distribution analysis using the Agilent 4200 (G2991AA) TapeStation system (Agilent, America).

**Translation following stress removal assay.** The highly enriched mRNA *glpK* in the aggresome was selected to perform the experiment. The pBAD vector was used as a backbone to generate a recombinant plasmid that expressing the GlpK-GFP fusion protein. Overnight bacterial cultures expressing pBAD-GlpK-GFP were 1:100 diluted into fresh LB medium and incubated at 37 °C with shaking at 220 rpm for 2 hr. 0.001% arabinose was then added to induce protein and mRNA expression of GlpK-GFP with continuous incubation for another 2 hr. Cell cultures were then treated with 2mM $NaAsO_2$ and 100ng/μL Rifampicin for 30 min. The transcription of *gplK*-mRNA in cytosol was inhibited by rifampicin, quenching GlpK-GFP fluorescence with laser at 488 nm. The fluorescence recovery rate of GlpK-GFP was quantified by continuously incubating the cells at 30 °C under a Nikon N-SIM S super-resolution microscope.

**Electron microscopy.** The cells were collected by centrifugation, and TEM fixative was added to the tube for fixation at 4°C. Agarose pre-embedding was performed using a 1% agarose solution, and the precipitation was suspended with forceps and wrapped in agarose before it solidified. The samples were then post-fixed by treating the agarose blocks with 1% $OsO_4$ in 0.1 M PB (pH 7.4) for 2 hr at room temperature. Subsequently, the samples were rinsed in 0.1 M PB (pH 7.4) three times for 15 mins each. To dehydrate the samples, a series of ethanol solutions were used at room temperature, followed by two changes of acetone. The samples were then incubated in pure EMBed 812 for 5-8 hr at 37°C. Polymerization was achieved by placing the samples in a 65 °C oven for more than 48 hr. Next, the resin blocks were cut into 50 nm thin sections using an ultramicrotome, and the tissue sections were transferred onto 150 mesh cuprum grids with formvar film. The sections were stained with a 2% uranium acetate saturated alcohol solution, avoiding light, for 8 mins, followed by staining in a 2.6% lead citrate solution, avoiding $CO_2$, for 8 mins. Finally, the cuprum grids were observed using a transmission electron microscope (Hitachi, HT7800/HT7700).

**Statistical analysis.** p > 0.05 was considered not significant. *P ≤ 0.05, **P ≤ 0.01, ***P ≤ 0.001, ****P ≤ 0.0001 by two-tailed Student's t test, one-way ANOVA, or Log-rank (Mantel-Cox) test. Statistical analysis was performed in GraphPad Prism or Excel.

**Slimfield imaging of Pepper RNA: HBC530 complex.** Slimfield[19] is an advanced light microscopy method which enables single molecule fluorescence detection and tracking to a lateral precision of approximately 40 nm[31] imaging over millisecond timescales, both *in vitro* and *in vivo*. The illumination field is generated by underfilling the back aperture of a high numerical aperture (NA) objective (1.49 in this case) with collimated laser light resulting in a delimited excitation volume large enough to encompass a number of whole bacterial cells at the sample plane whose excitation intensity is greater than that of traditional epifluorescence microscopy by over three orders of magnitude. The excitation of fluorescent proteins both in single- and dual-colour bacterial cell strains enables a typical exposure time of 1-5 ms per frame[32-34], down to a few hundred microseconds when using bright organic dyes[35,36]. By using bespoke single-particle tracking software[37,38] and utilising step-wise photobleaching of dyes[39,40] and edge-preserving filtration[41], the diffusion coefficient and the stoichiometry of tracked molecules and molecular assemblies can be accurately determined, rendering distributions via kernel density estimation[42].

For *in vitro* imaging or surface-immobilized mRNA, Polyethelene glycol (PEG) passivated slides were prepared similarly to Paul and Myong[43] except that MeO-PEG-NHS and Biotin-PEG-NHS (Iris biotech PEG1165 and PEG1057 respectively) were used for passivation. Flow cells were made by attaching a PEG passivated coverslip to a PEG passivated slide via two pieces of double-sided tape with a 5 mm gap between them. The flow cell was first incubated for 5 mins at room temperature with 200 μg/mL neutravidin (Thermofisher Scientific, 31000) in Phosphate Buffered Saline (PBS) followed by a 200 μL wash with PBS to remove excess neutravidin. 100 μL of 30 pM Pepper RNA in PBS was then introduced into the to the flow cell and incubated for 10 mins at room temperature to allow binding of the 5' biotin on the RNA molecule to the neutravidin bound to the flow cell surface, excess PEPPER RNA was washed off with a 200 μL PBS wash. Finally, 200 μL of imaging buffer (40 mM HEPES, pH 7.4, 100 mM KCl, 5mM $MgCl_2$, 1x HBC530) was washed into the flow cell before the open channel ends were sealed. A bespoke single-molecule microscope[32] was utilised in order to image the surface immobilised Pepper RNA:HBC530 complex[1]. Excitation by an Obis LS 50 mW 488 nm wavelength laser (attenuated to 20 mW) was reduced to 10 μm at full width half maxima in the sample plane to produce a narrow field of illumination[44], with a mean excitation intensity of 0.25 mW/μm². Images were magnified to 53nm/pixel and captured using a prime 95B sCMOS camera with a 5 ms exposure time. Image analysis was carried out on 60 fields of view from three separate slide preparations using previously described bespoke MATLAB scripts[35].

For *in vivo* imaging, cells were grown overnight in LB at 37 °C with shaking at 180 rpm. The overnight cultures were diluted 1:100 in fresh LB and grown at 37 °C for 2 hr with shaking at 180 rpm. Arabinose was then added at a final concentration of 0.001% and the cultures were incubated for another 4 hr at 22 °C with shaking at 220 rpm. The cultures were then centrifuged at 4000g for 5 min to collect the cells. The collected cells were resuspended in imaging buffer with HBC ligand (1:100 dilution) and 0.001% arabinose and incubated at room temperature for 5-15 min. The cells were then spotted on agarose gel pads containing M9 medium with HBC ligand (1:100), 25 mM HEPES (pH 7.4), 5 mM $MgSO_4$, 20mM arsenite and 0.01% arabinose. The cells were then imaged on a Slimfield microscope set up using the same imaging settings as for *in vitro* imaging and analysed using the same tracking software.

**Modeling.** Aggresome formation was interpreted as LLPS involving a ternary mixture of protein, mRNA and aqueous solvent, driven by net gain in attractive enthalpy mediated from increased protein and RNA interactions which outweigh the loss in entropy due to a more ordered phase-separated state in the aggresome compared to solvated protein and RNA in the cytosol. The enthalpic component embodied in the χ interaction parameter depicts overall enthalpy as the sum of interactions involving the water solvent, protein, and RNA while the entropy component characterizes biomolecular mixing and translational effects. In the limit of low RNA concentrations, as suggested from our earlier measurements indicating a protein-rich aggresome environment[2], χ is dominated by protein-solvent interactions. We assume that in the low RNA concentration limit RNA can partition into aggresomes but has negligible influence on phase separation. The enthalpic component of RNA-protein and RNA-water interactions per nucleotide base are assumed to be non-specific. For modeling RNA degradation effects, RNA diffusion in the cytosol was modeled using the Stokes-Einstein relation assuming a length-dependent radius of gyration as predicted by a freely-jointed chain. For simplicity, RNA degradation was modeled with a fixed probability density at each nucleotide base. The probability of RNA escape from an aggresome once incorporated was modeled as a Bolztmann factor incorporating a length-dependent enthalpic factor. Full details in Supplementary Information.

**Protein surface charge analysis.** For the surface charge analysis of proteins mentioned in this study, the structure of relative proteins was downloaded from Uniprot in AlphaFold format and imported into PyMOL

([http://www.pymol.org](http://www.pymol.org)), a user-sponsored molecular visualization system on an open-source foundation, to calculate surface charge using the vacuum electrostatics generation function. For the mutants, surface charges were analyzed based on wild-type proteins. Relative amino acids were mutated using the mutagenesis function of PyMoL and recalculated to obtain the surface charges of mutants.

**Bioinformatics analysis.** For the RNA-Seq analysis, the raw data quality of the prepared libraries was assessed using FastQC (http://www.bioinformatics.babraham.ac.uk/projects/fastqc/). The reads were mapped to the MG1655 k12 genome (Ensembl Bacteria, Taxonomy ID: 511145) using the BWA aligner software (version 0.7.17-r1188, https://github.com/lh3/bwa.git). Then the sam files was converted to bam files using samtool (Version 1.9, https://samtools.sourceforge.net/). The bam files were counted with featureCounts (Version 2.0.1, https://github.com/topics/featurecounts) to generate expression results. Differential expression analysis was performed using DESeq2[45]. For Gene Ontology (GO) analysis and Kyoto Encyclopedia of Genes and Genomes (KEGG) pathway analysis, gene id was converted using the DAVID web resource (http://david.abcc.ncifcrf.gov/). R package clusterProfiler R package[46] was used to identify the pathways in which these up- and down-regulate genes are enriched.

For the mass spectrometry analysis, aggresome and cytosol proteins were analyzed using mass spectrometry (MS), and the resulting data were first normalized by the total number of proteins per sample. Then, differential analysis was performed using DESeq2 to find the differentially expressed proteins. Gene Ontology (GO) analysis was performed using clusterProfiler R package for the enriched proteins of the samples. Protein-protein interaction network functional enrichment was then analyzed using STRING[47] (https://cn.string-db.org/) for proteins enriched in aggresome or cytosol with an average expression abundance greater than 50. In addition to this, the GeneMANIA prediction server (biological network integration for gene prioritization and predicting gene function. http://pages.genemania.org/) was used to analyze the proteins. The interaction data were exported and later plotted using Cytoscape (An open-source platform for complex analysis and visualization. https://cytoscape.org/).

**Supplementary figures**

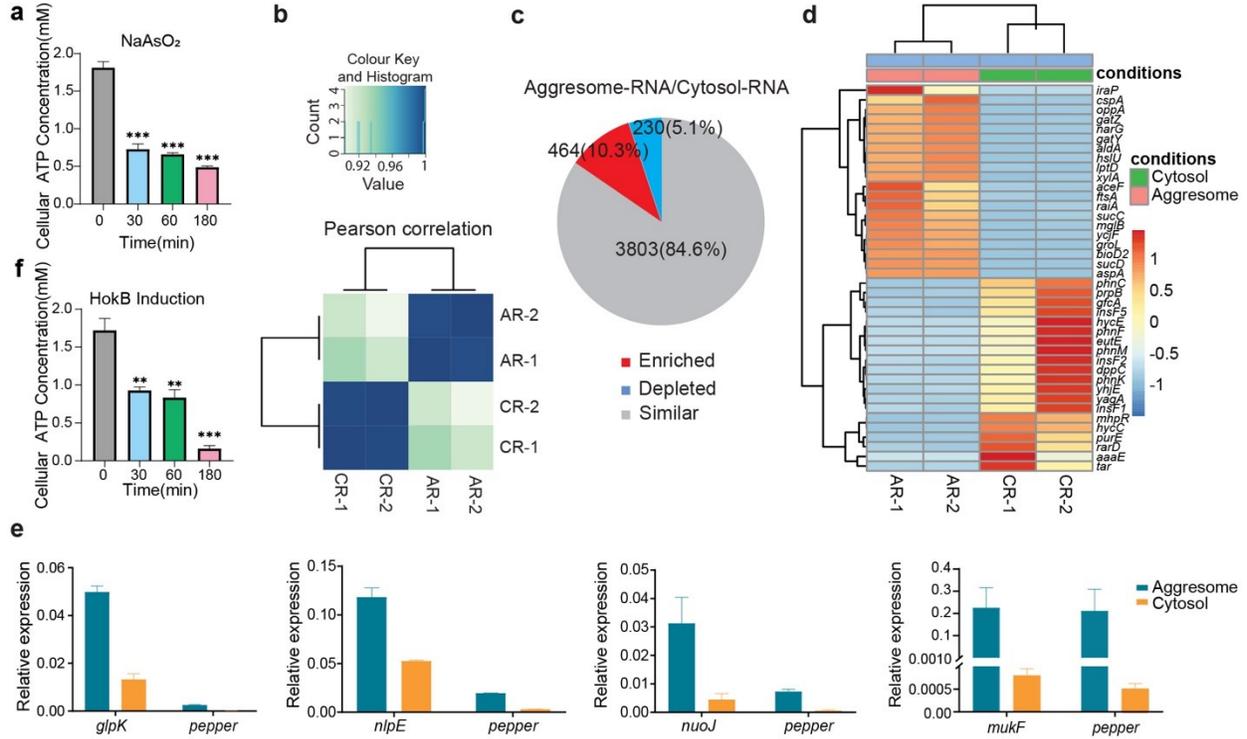

Extended Data Fig. 1 | Aggresome formation enriches mRNA. **a**. Cellular ATP concentration after arsenite (2mM) treatment for various time durations. **b**. Pairwise correlation coefficients between Aggresome-RNA library duplicates and Cytosol-RNA library duplicates, indicating that the aggresome transcriptome is distinct from that of the cytosol (Pearson correlation coefficient, $R_2 < 0.001$). **c**. Pie chart depicting gene number and the relative contribution of each class of RNA (Aggresome enriched, Aggresome depleted, or neither) to the cytosol transcriptome. **d**. Heatmap showing relative transcript abundance of Aggresome-RNA and Cytosol-RNA. Scale beside the heatmap indicates log2-normalized transcript abundance relative to the mean expression level. **e**. mRNA expression level of the genes measured by quantitative RT-PCR. **f**. Cellular ATP concentration during HokB induction (0.001% arabinose) for various time durations. Unpaired Student's t-test used; error bars indicate SE; *P <0.05, **P < 0.005, and ***P < 0.0005.

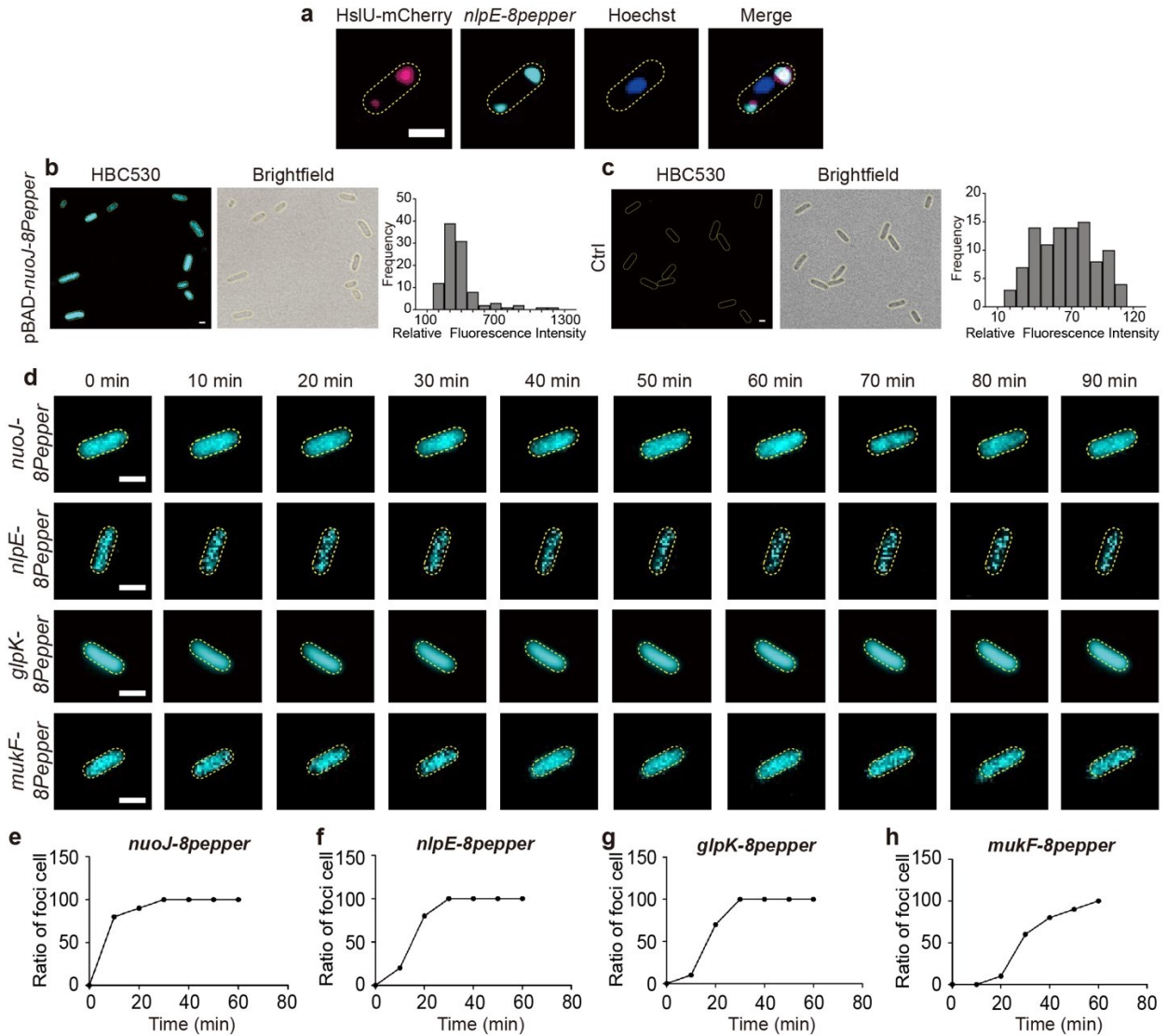

Extended Data Fig. 2 | **a**. Epifluorescence imaging showing the macromolecule composition of the bacterial aggresome (induced by 2 mM arsenite, 30 min). *nlpE-8pepper*, RNA aptamer/HBC labeled *nlpE* mRNA; Hoechst: DNA labeling. **b**. Epifluorescence imaging of live *E. coli* cells expressing Pepper aptamer, stained with HBC530 dye (1 μM). **c**. Epifluorescence imaging of live wild-type *E. coli* cells, stained with HBC530 dye (1 μM). **d**. Distribution of *glpK-8pepper* mRNA in cells without stress stimuli. **e-h**. Proportion of cells showing fluorescent foci as a function of arsenite treatment time, *nuoJ* (**e**), *nlpE* (**f**), *glpK* (**g**) and *mukF* (**h**) (n=10). Scale bar, 1 μm.

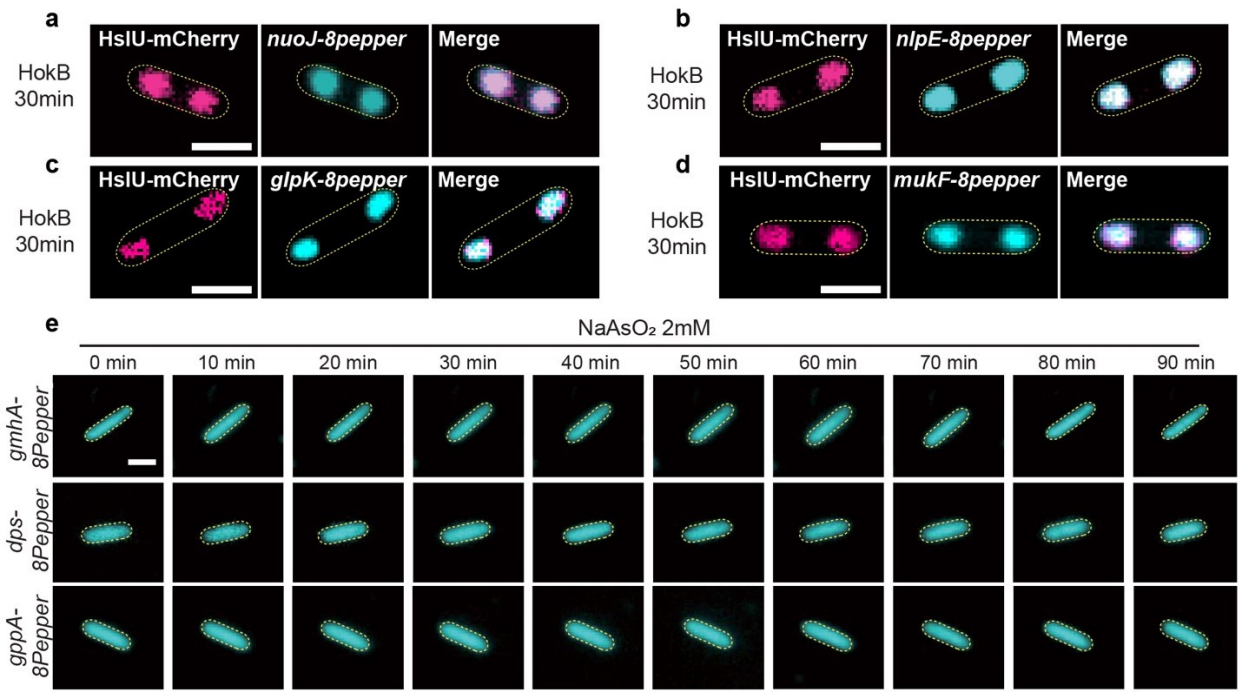

Extended Data Fig. 3 | **a-d**. SIM imaging showing that mRNAs of *nuoJ* (**a**), *nlpE* (**b**), *glpK* (**c**) and *mukF* (**d**) partition into aggresomes under HokB induction for 30 min. **e**. Distribution of *gmhA/dps/gppA-8pepper* mRNAs in cells with arsenite treatment. mRNAs are labeled by 8pepper and imaged in the presence of HBC dye in imaging buffer. Scale bar, 2 μm.

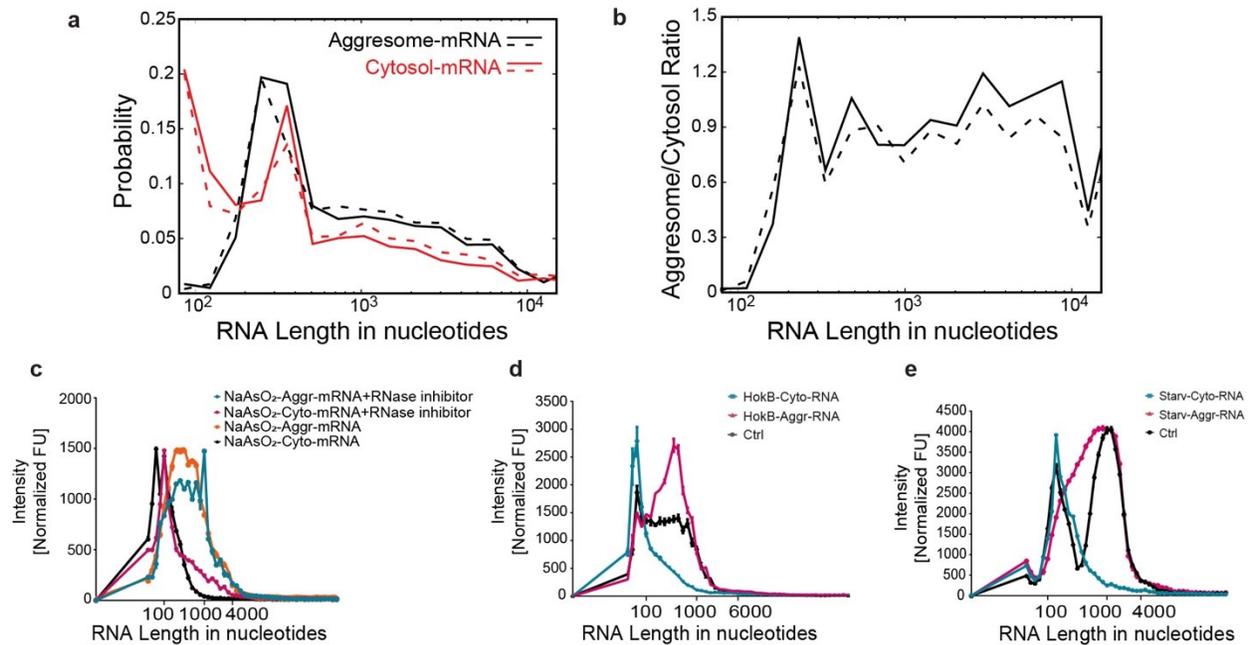

Extended Data Fig. 4 | **a**. Distribution of RNA length in aggresomes and cytoplasm, analyzed by using transcriptome sequencing data with operonic mRNA as a reference. **b**. The ratio between aggresome RNA and cytoplasm RNA as a function of RNA length in nucleotides, analyzed by using transcriptome sequencing data with operonic mRNA as a reference. **c**. ScreenTape RNA analysis of Aggresome-mRNA (Ars-AmR) and Cytosol-mRNA (Ars-CmR) from the cells with arsenite treatment for 30 min (n=3), with or without RNase inhibitor during mRNA enrichment. **d**. ScreenTape RNA analysis of Aggresome-RNA (Aggr-RNA) and Cytosol-RNA (Cyto-RNA) from the cells with HokB induction for 30 min, and total RNA content from the cells grown in the exponential phase is used as the control (Ctrl) (n=3). **e**. ScreenTape RNA analysis of Aggresome-RNA (Aggr-RNA) and Cytosol-RNA (Cyto-RNA) from the cells with starvation (Starv) induction for 24 hours, and total RNA content from the cells grown in the exponential phase is used as the control (Ctrl) (n=3; error bar indicates SE).

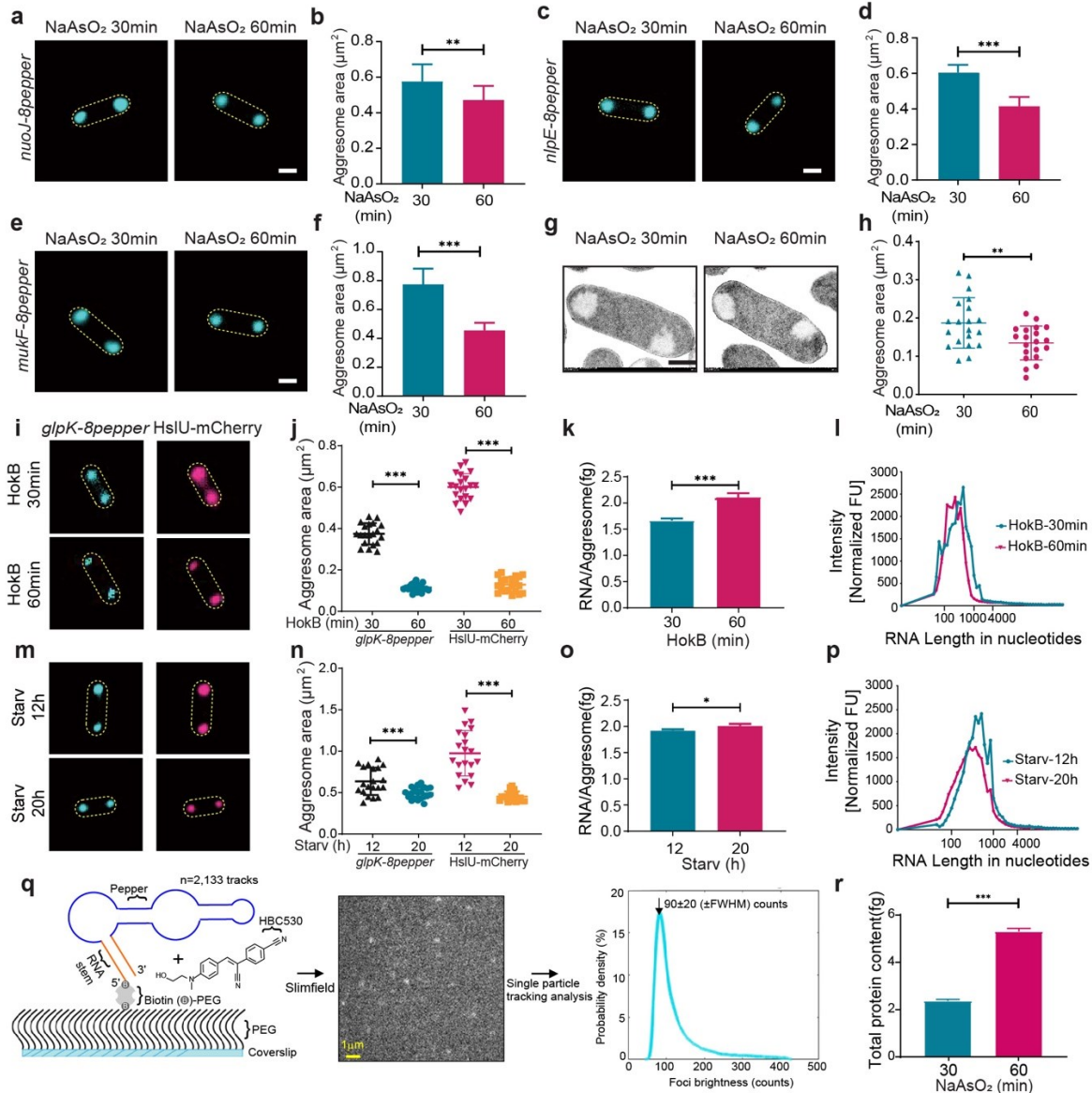

Extended Data Fig. 5 | **a-f**. SIM images of aggresome RNA compaction. **a, c, e**. Representative SIM images of cells expressing *nuoJ-8pepper* (**a**), *nlpE-8pepper* (**c**) and *mukF-8pepper* (**e**) with different duration of arsenite (Ars) treatment, images were taken in the presence of HBC530 dye, Scale bar, 1 μm. **b, d, f**. Statistical analysis of aggresome area from SIM images from (**a, c, e**), respectively (n=10). Error bar indicates SE; *P <0.05, **P < 0.005, and ***P < 0.0005. **g**. Representative TEM images of cells with different duration of arsenite (Ars) treatment. Scale bar, 500 nm. **h**. Statistical analysis of aggresome area in cells with different duration of arsenite treatment, based on TEM images (n=20). **i, m**. Representative SIM images of cells with different duration of HokB treatment (**i**) or starvation (Starv) (**m**), *glpK-8pepper* represents *glpK-8pepper* RNA stained with HBC530 green dye, HslU-mCherry represents protein HslU labeled by fluorescent protein mCherry. **j, n**. Statistical analysis of aggresome area in cells with different duration of HokB treatment or starvation (Starv), based on SIM images from (**i, m**) (n=20). **k, o**. The mass of RNA per aggresome in cells with different duration of HokB treatment or starvation (Starv), the total mass of RNA in bulk aggresomes (RNA$_{total}$) is measured by a Qubit fluorometer, the number of aggresomes (N$_{aggresome}$) is determined by FACS, RNA/Aggresome is determined by RNA$_{total}$/N$_{aggresome}$ (n=3). **l, p**. Distribution of RNA length in aggresomes determined by ScreenTape RNA analysis, aggresomes from cells with different durations of HokB treatment or starvation (Starv) (n=3). **q**. Pipeline for in vitro single-molecule mRNA detection. (Left panel) Pepper RNA construct is immobilized to a passivated coverslip surface via the 'stem' (orange); the Pepper aptamer (blue) is specific for an organic dye, which is then incubated with it (here 'HBC530').

(Middle panel) high sensitivity millisecond timescale Slimfield localizes these bound dye molecules as diffraction limited fluorescent foci. (Right panel) These foci can then be tracked used automated home-written single-particle tracking analysis software which pinpoints the intensity centroid to within approximately 40 nm precision; the brightness of foci is calculated by summing all pixel intensities and subtracting a local background value, here indicating a modal value of single-molecule brightness of approximately 90 counts on our CMOS camera detector. **r**, The mass of protein per aggresome in cells with different duration of arsenite exposure, the total mass of protein in bulk aggresomes (Protein $_{total}$) is measured by Qubit fluorometer, the number of aggresomes ($N_{aggresome}$) is determined by FACS, Protein/Aggresome is determined by Protein $_{total}$/$N_{aggresome}$ (n=3). Unpaired Student's t test used; error bar indicates SE; *$P < 0.05$, **$P < 0.005$, and ***$P < 0.0005$.

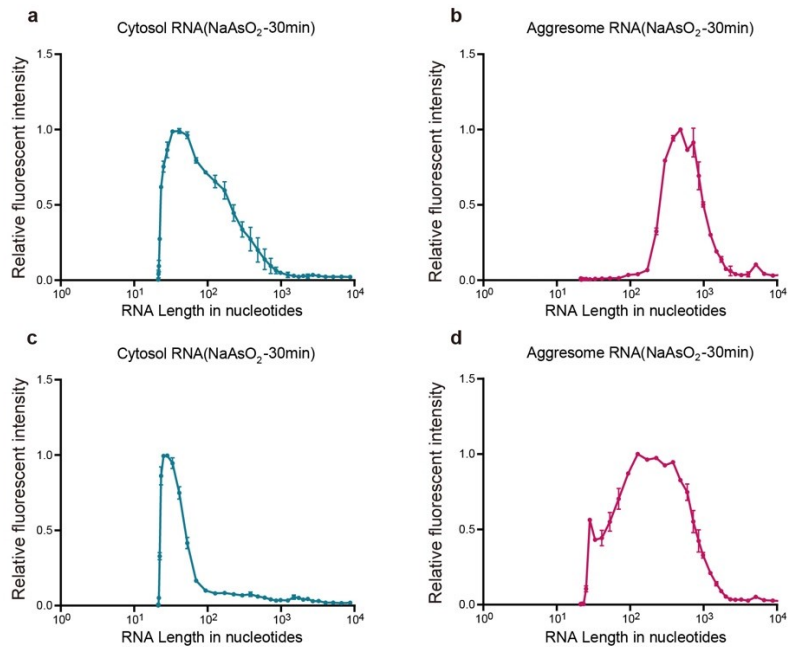

Extended Data Fig. 6 | ScreenTape RNA analysis of Aggresome-RNA (Aggr-RNA) and Cytosol-RNA (Cyto-RNA) from the cells with arsenite treatment for 30 min and 60 min. The fluorescence intensity was normalized using the Min-Max scaling method to standardize the data across samples (n=3; error bar indicates SE).

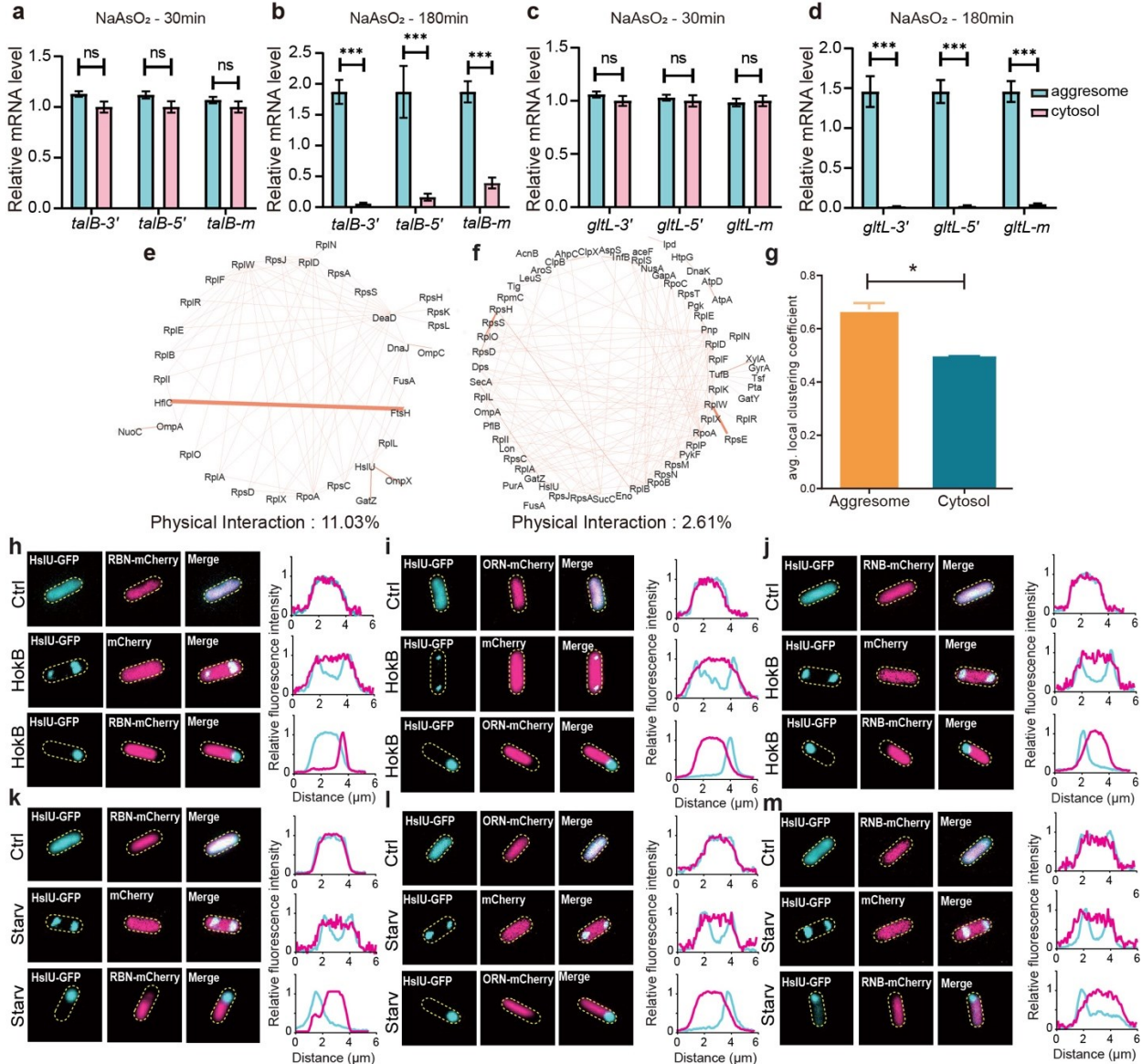

Extended Data Fig. 7 | **a-d**. Relative RNA expression level of *talB* and *gltL* in aggresomes and cytosol after 30 min or 180 min of 2 mM arsenite treatment (n=3). **e**. Protein-protein interaction (PPI) networks for aggresome proteins. **f**. PPI networks for cytoplasm proteins. **g**. Average local clustering coefficients for aggresome and cytoplasm proteins. **h-j**. Distribution of ribonucleases, RBN (**h**), ORN (**i**) or RNB (**j**), in cells with toxin protein HokB overexpression. **k-m**. Distribution of ribonucleases, RBN (**k**), ORN (**l**) or RNB (**m**), in starved cells. Aggreseome marker HslU is labeled by GFP. ribonucleases are labeled by mCherry. Relative fluorescence intensities of GFP and mCherry along the long axis of each cell are measured and plotted (chart on the right of each row of fluorescence images). Ctrl, control, indicating exponential phase cells without drug treatment; Ars, arsenite treatment. Scale bar, 1 μm. Unpaired Student's t-test was performed, and the error bars indicate SE; *$P <0.05$, **$P <0.005$, and ***$P < 0.0005$.

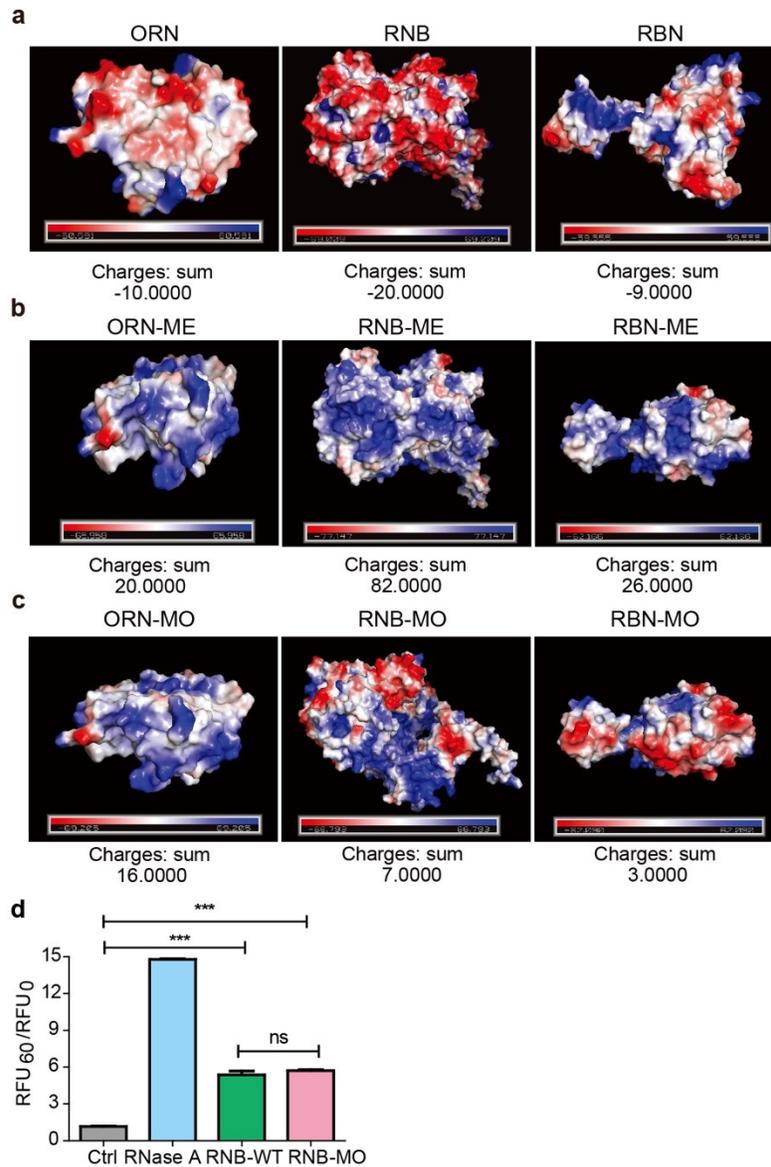

Extended Data Fig. 8 | Protein surface charge determined by PyMOL. **a**. Protein surface charges of wild-type ORN, RNB, RBN. **b**. Protein surface charges of mutated ribonucleases, by substituting all aspartic acid (D) or glutamic acid (E) residues across the entire enzymatic center with alanine (A) residues. ME: **M**utation across the **E**ntire enzymatic center. **c**. Protein surface charges of mutated ribonucleases, by substituting aspartic acid (D) and glutamic acid (E) residues with alanine (A) residues in the region outside DNA binding region and catalytic center. MO: **M**utation **O**utside the regions of RNA binding motif and catalytic center. **d**. RNase enzyme activity measurement. RNase A serves as the positive control, while "Ctrl" denotes the buffer-only control. Statistical analysis was conducted using an unpaired Student's t-test, with error bars representing the standard error (SE). Asterisks denote significance levels as follows: *P < 0.05, **P < 0.005, and ***P < 0.0005.

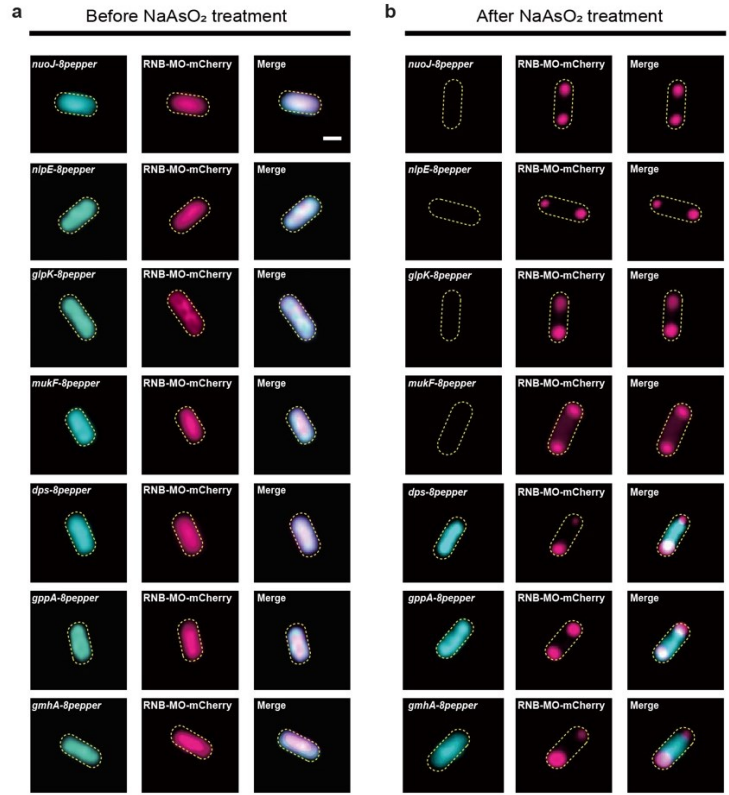

Extended Data Fig. 9 | **a**. SIM imaging showing that mRNAs of *nuoJ/nlpE/glpK/mukF/gmhA/dps/gppA-8pepper* mRNAs in cells without stimuli. **b**. SIM imaging showing that mRNAs of *nuoJ/nlpE/glpK/mukF/gmhA/dps/gppA-8pepper* mRNAs in cells with 2mM arsenite treatment for 30 min. mRNAs are labeled by an 8-pepper tag and imaged in the presence of HBC dye in imaging buffer. Scale bar, 1 μm.

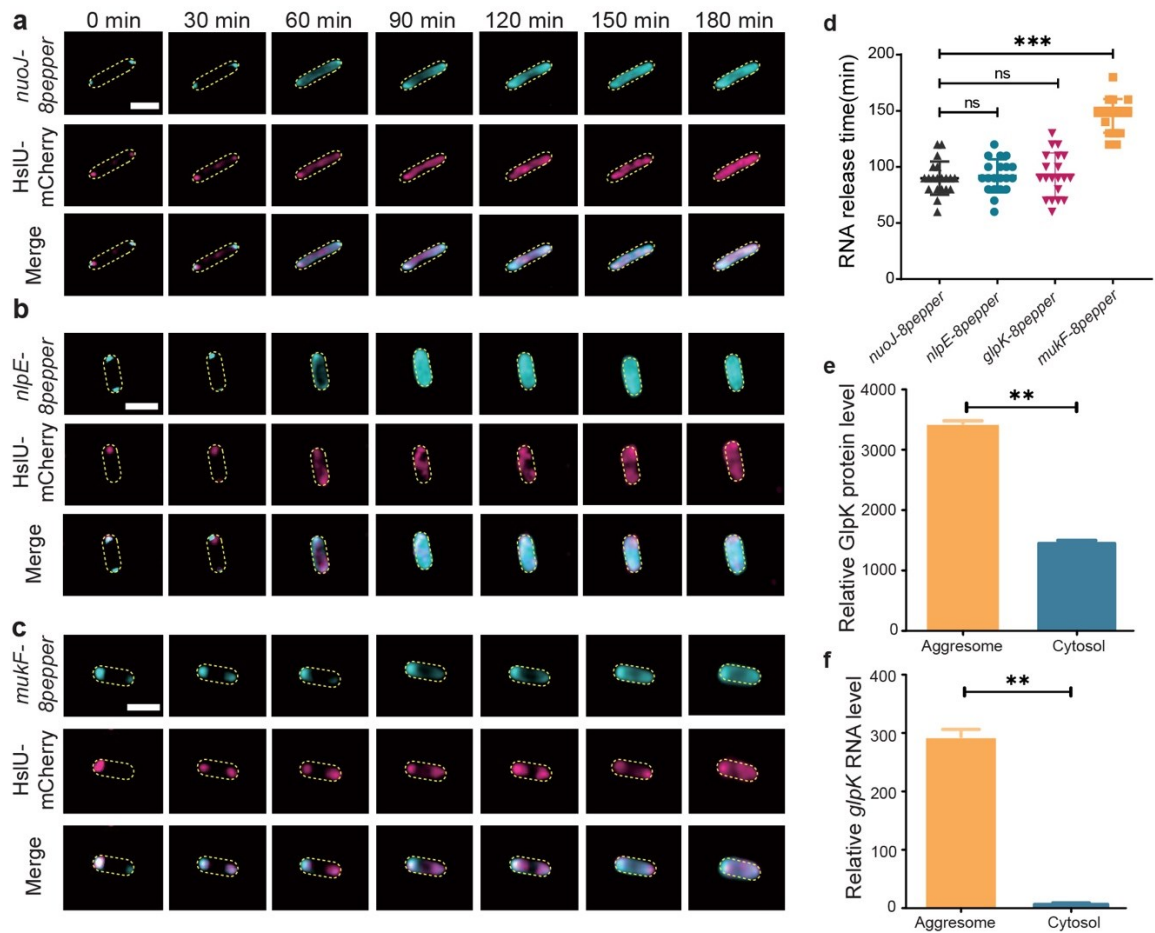

Extended Data Fig. 10 | **a-c**. Releasing of *nuoJ* (**a**), *nlpE* (**b**) and *mukF* (**c**) mRNA from aggresomes after removal of arsenite. Aggresome marker HslU is labeled by mCherry. mRNAs are labeled by 8pepper and imaged in the presence of HBC dye in imaging buffer. Scale bar, 2 μm. **d**. The average time for each type of mRNA releasing (n=20). **e**. Relative GlpK protein levels in aggresome and cytoplasm after arsenite treatment, measured by MS (n=4). **f**. Relative *glpK* RNA levels in aggresome and cytoplasm after arsenite treatment, measured by RNA-seq (n=10). Error bars indicate SE; *P <0.05, **P < 0.005, and ***P < 0.0005.

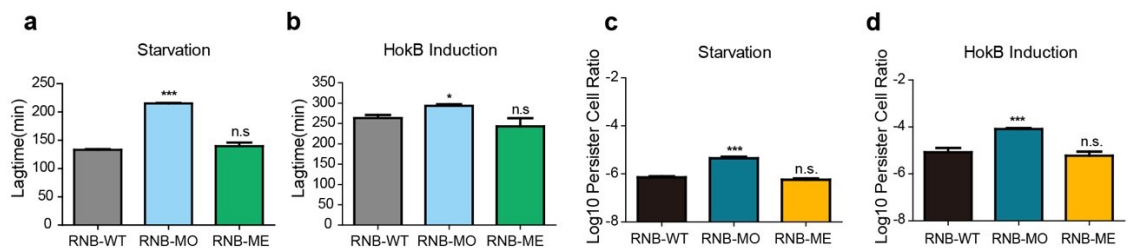

Extended Data Fig. 11 | Phenotypic consequences of RNA degradation in aggresomes. **a-b**. Lag time of strains containing wild-type RNB or RNB-MO and RNB-ME after exposure to starvation for 24 hours (**a**) or HokB expression with 0.001% arabinose (w/v) for 30 minutes (**b**). **c-d**. Persister cell ratio in strains containing wild-type RNB or the mutant RNB-MO and RNB-ME after exposure to after exposure to starvation for 24 hours (**c**) or HokB expression with 0.001% arabinose (w/v) for 30 minutes (**d**). Statistical analysis was performed using an unpaired

Student's t-test, with error bars indicating the standard error (SE). Asterisks denote significance levels as follows: *$P < 0.05$, **$P < 0.005$, and ***$P < 0.0005$.

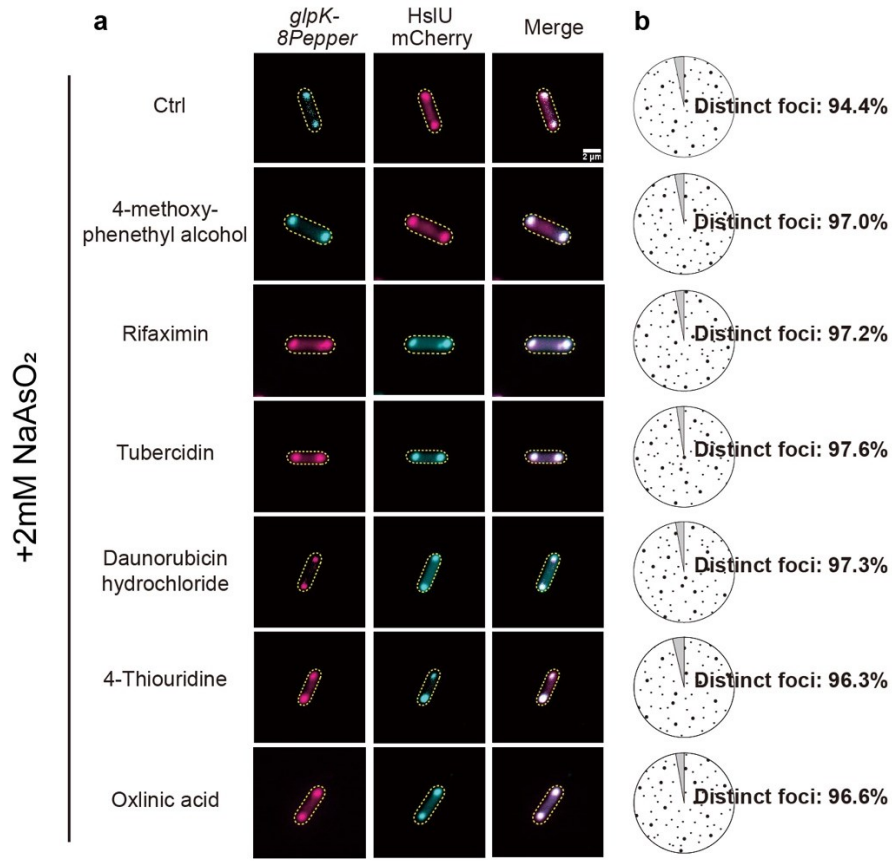

Extended Data Fig. 12 | Small chemical compounds screening to identify chemicals that disrupt RNA recruitment into aggresomes. **a**. Representative fluorescence images of cells with different chemical combination treatment. **b**. Statistical analysis of cells with distinct mRNA foci after different chemical combination treatment, data from (**a**) (n=50, 3 replicates).

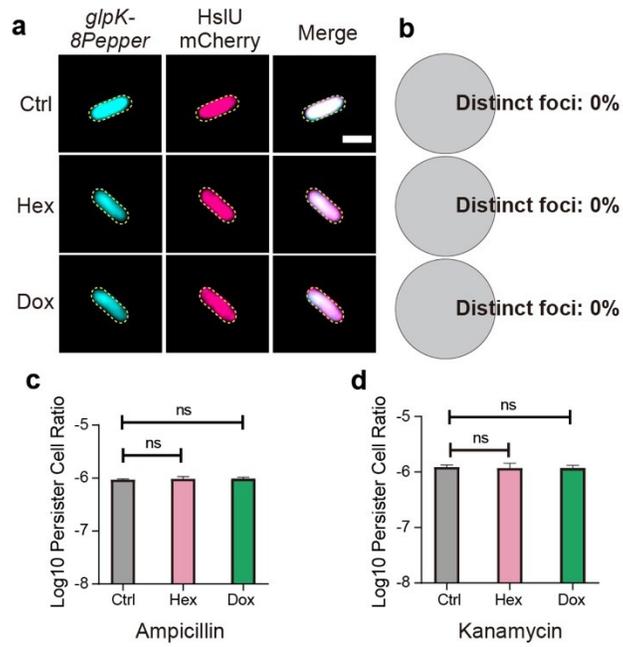

Extended Data Fig. 13| Evaluation of the effects of Hex and Dox on cell survival. **a**. Representative fluorescence images of cells with different chemical treatment. **b**. Statistical analysis of cells with distinct mRNA foci after different chemical combination treatment, data from (**a**) (n=100). **c-d**. Cell survival rate (log scale) after 4 hours **(c)**ampicillin **(d)**kanamycin killing (n=3). Ctrl, control; Dox, doxorubicin; Hex, 1,6-hexanediol. Scale bar, 1 μm. Unpaired Student's t-test was performed, and the error bars indicate SE; n=3, *$P < 0.05$, **$P < 0.005$, and ***$P < 0.0005$.